\documentclass[a4paper,fleqn,usenatbib]{mnras}
\usepackage{amsmath}
\usepackage{txfonts}

\usepackage[T1]{fontenc}
\usepackage{ae,aecompl}

\usepackage{graphicx}	
\usepackage{grffile}

\usepackage{epsf}

\AtBeginDocument{}%
\begin{document}
\newcommand{\msun}{M_{\odot}}
\newcommand{\rsun}{M_{\odot}}
\newcommand{\kms}{\, {\rm km\, s}^{-1}}
\newcommand{\cm}{\, {\rm cm}}
\newcommand{\gm}{\, {\rm g}}
\newcommand{\erg}{\, {\rm erg}}
\newcommand{\kel}{\, {\rm K}}
\newcommand{\kpc}{\, {\rm kpc}}
\newcommand{\mpc}{\, {\rm Mpc}}
\newcommand{\seg}{\, {\rm s}}
\newcommand{\kev}{\, {\rm keV}}
\newcommand{\hz}{\, {\rm Hz}}
\newcommand{\etal}{et al.\ }
\newcommand{\yr}{\, {\rm yr}}
\newcommand{\gyr}{\, {\rm Gyr}}
\newcommand{\eq}{eq.\ }
\newcommand{\amunit}{\msun {\rm AU^2/yr}}
\def\arcsec{''\hskip-3pt .}

\def\gapprox{\;\rlap{\lower 3.0pt                       
             \hbox{$\sim$}}\raise 2.5pt\hbox{$>$}\;}
\def\lapprox{\;\rlap{\lower 3.1pt                       
             \hbox{$\sim$}}\raise 2.7pt\hbox{$<$}\;}

\newcommand{\figsizeFour}{9.0cm}

\newcommand{\figwidthSingle}{14.0cm}
\newcommand{\figwidthDouble}{7.50cm}

\newcommand{\figbig}{\figwidthSingle}
\newcommand{\figsmall}{\figwidthDouble}

\newcommand{\figappend}{\figwidthSingle}

\newcommand{\reqOne}[1]{equation~(\ref{#1})}
\newcommand{\reqTwo}[2]{equations~(\ref{#1}) and~(\ref{#2})}
\newcommand{\reqNP}[1]{equation~\ref{#1}}
\newcommand{\reqTwoNP}[2]{equations~\ref{#1} and~\ref{#2}}
\newcommand{\reqTo}[2]{equation~(\ref{#1})-(\ref{#2})}
\newcommand{\rn}[1]{(\ref{#1})}
\newcommand{\ern}[1]{equation~(\ref{#1})}
\newcommand{\be}{\begin{equation}}
\newcommand{\ee}{\end{equation}}
\newcommand{\ff}[2]{{\textstyle \frac{#1}{#2}}}
\newcommand{\ben}{\begin{enumerate}}
\newcommand{\een}{\end{enumerate}}

\title[Star--disc interaction in galactic nuclei]
{Star--disc interaction in galactic nuclei: formation of a central stellar disc}

\author[Panamarev et al]
  {Taras Panamarev$^{1,2,3}$\thanks{Corresponding author email: taras@ari.uni-heidelberg.de}\thanks{Fellow of the International Max Planck Research School for Astronomy
  		and Cosmic Physics at the University of Heidelberg (IMPRS-HD)}, Bekdaulet Shukirgaliyev$^{1,2,3}$\footnotemark[2], Yohai Meiron$^{4,3}$,\\
   \newauthor Peter Berczik$^{3,5,1}$,  Andreas Just$^1$, Rainer Spurzem$^{3,6,1}$, \\
   \newauthor Chingis Omarov$^2$, Emmanuil Vilkoviskij$^2$\\
   \\
      $^1$ Zentrum f\"ur Astronomie der Universit\"at Heidelberg, Astronomisches Rechen-Institut, M\"onchhofstr. 12-14, 69120 Heidelberg, Germany \\
      $^2$ Fesenkov Astrophysical Institute, Observatory 23, 050020 Almaty, Kazakhstan \\
      $^3$ National Astronomical Observatories of China, Chinese Academy of Sciences, 20A Datun Rd., Chaoyang District, 100012, Beijing, China\\
      $^4$ Institute of Physics, E\"otv\"os University, P\'azm\'any P.s. 1/A, Budapest 1117, Hungary\\
      $^5$ Main Astronomical Observatory, National Academy of Sciences of Ukraine, 27 Akademika Zabolotnoho St., 03680, Kyiv, Ukraine\\
      $^6$ Kavli Institute for Astronomy and Astrophysics at Peking University, 5 Yiheyuan Rd., Haidian District, 100871, Beijing, China
}

\maketitle

\begin{abstract}

We perform high resolution direct $N$-body simulations to study the effect of an accretion disc on stellar dynamics in an active galactic nucleus (AGN). We show that the interaction of the nuclear stellar cluster (NSC) with the gaseous disc (AD) leads to formation of a stellar disc in the central part of the NSC. The accretion of stars from the stellar disc onto the super-massive black hole is balanced by the capture of stars from the NSC into the stellar disc, yielding a stationary density profile. We derive the migration time through the AD to be 3\% of the half-mass relaxation time of the NSC. The mass and size of the stellar disc are 0.7\% of the mass and 5\% of the influence radius of the super-massive black hole. 
An AD lifetime shorter than the migration time would result in a less massive nuclear stellar disc. The detection of such a stellar disc could point to past activity of the hosting galactic nucleus.

\end{abstract}


\section{Introduction}\label{sec:INTRO}

The early universe features extremely luminous objects known as quasars. A quasar represents the central region of an active galaxy that contains a super-massive black hole (SMBH) with mass $M_\mathrm{SMBH} = 10^7-10^{10} \msun$ surrounded by an accretion disc (AD). As gas infalls from large scales it settles into a disc-like structure initially losing energy and transporting angular momentum outwards. It is still debated in the literature how exactly the central black holes reach such extreme masses. The activity of quasars peaks around a redshift of $z=2$ and makes it even more difficult to explain the growth mechanism of the SMBH. While quasars are the most luminous objects known in the universe, there are other types of active galactic nuclei such as Seyfert galaxies, blazars and radio galaxies. All these objects are unified under the assumption of being the same type of astrophysical object observed from different angles \citep{Antonucci1993,Urry1995} however this simple unification scheme is criticised \citep[e.g.][for a review]{Netzer2015}.  

In order to explain high luminosities, an accretion disc is required. One of the first analytical descriptions of the AD goes back to \citet{SS1973} and \citet{NT1973}, where they derived AD density profile for the case of stellar black holes in binary systems where a black hole is fed by the secondary component. More recent models of accretion discs are extensions of the Shakura \& Sunyaev thin disc model. 

Besides the SMBH and the AD, a compact stellar cluster is another main component of an AGN. The interaction of stars with the AD was studied by \citet{Rauch1995,Rauch1999}. Later a semi-analytic approach of star--disc interactions \citep{VilkoviskijCzerny2002} led to the conclusion that the competition between stellar two-body  relaxation and dissipation will cause a disc-like structure in the inner stellar component and a well-defined stationary flux of stars towards the SMBH. After the material from the gaseous disc is used up, the AGN goes over to a quiescent phase.

The Milky Way (MW) galaxy is among the quiescent ones, but it is debated whether it has been active in the past \citep[e.g.][]{Nayakshin2005}. The Galactic centre hosts a SMBH with $M_\mathrm{SMBH} = 4.3\times 10^6 \msun$ \citep[see][for the latest value]{Gillessen2017} and a nuclear star cluster (NSC). High resolution adaptive optic observations of the central parsec revealed a cluster of young ($< 100$ Myr old, spectral class B) high velocity stars inside the inner arcsec (0.04 pc). Moreover, a disc of even younger stars ($6\pm 2$ Myr old, O/Wolf-Rayet stars) was discovered at a distance between 0.04 and 0.5 pc from SgrA* \citep{LevinBeloborodov2003}. The disc appears to be highly warped \citep{Bartko2009,Bartko2010}. The analysis of high quality proper motion measurements of bright massive stars suggests the existence of a second stellar disc with opposite rotation \citep[][for a review]{PaumardEtAl2006,Genzel2010}, however the stellar distribution may be more irregular \citep{Kocsis2011}. The most prominent disc exhibits clockwise rotation and its surface density scales as $\Sigma \propto R^{-\gamma}$ with $\gamma$ ranging from $\simeq 1.4 \pm 0.2$ \citep{Bartko2010} to $ \simeq 2.3\pm 0.7$ \citep{Lu2009}. The disc(s) is located inside a massive spherical cluster of old stars, which is two orders of magnitude more massive than the disc \citep{PaumardEtAl2006,Bartko2010,Yelda2014}. One possible explanation for the origin of the stellar disc(s) is that a precursor gaseous accretion disc underwent a fragmentation phase a few million years ago leading to the formation of new stars \citep{Nayakshin2005,Nayakshin2007}. \citet{Amaro-Seoane2014} and \citet{Kieffer2016} showed that collisions with the fragmenting clumps may destroy the outer envelopes of red giants leading to shallow or even flat density profile of old stars (\citealt{Buchholz2009}, but see \citealt{Gallego-Cano2017}).

The effects of gas damping in dense stellar systems were studied by \citet{Leigh2014} analytically and numerically. The authors conclude that the gas drag may increase the stellar accretion rate onto the SMBH in galactic nuclei while the effect of the star-gas interactions on the mass segregation rate is relatively inefficient in case of dense galactic nuclei. Stellar migration towards the SMBH in AGN was analysed by \cite{Mckernan2011} where the authors considered compact massive stellar objects migrating by analogy to protoplanetary migration. In result, the migration and accretion of compact objects can explain the X-ray soft excess in Seyfert AGN. \citet{Baruteau2011} performed hydrodynamical simulations of the gaseous disc in order to study the migration of a binary star through the disc and they found that the hardening of the binary happens on much shorter time-scales than the migration towards the SMBH. 

It is natural to expect the presence of stellar mass black holes (sBH) in the AD where they can accrete material and grow or even accumulate in a migration trap and merge resulting in the formation of an intermediate mass black hole \citep{ArtymowiczEtAl1993,Bellovary2016}. The gaseous drag would effectively reduce the semi-major axis of sBH binaries resulting in a strong gravitational wave emission followed by a merger within the lifetime of the AD \citep{Bartos2017, Stone2017,McKernan2017}. 

This paper is an extension of \citet{JustEtAl2012} (hereafter Paper~I) and \citet{KenEtAl2016} (hereafter Paper~II). In Paper~I we examined the accretion rate onto the SMBH and found that the presence of the AD enhances the accretion while the focus of Paper~II was on the orbital parameters of accreted stars leading to the conclusion that the stars trapped by the gas accrete with near-circular orbits. \citet{Shukirgaliyev2016} noted the presence of a stellar disc-like structure in the inner part of the NSC. In this paper we study the properties of the nuclear stellar disc (NSD) in AGN using the most realistic simulation from Paper~II with $N = 1.28\times10^5$ particles. 

Method for modelling star--star and star--gas interactions as well as the initial conditions for the stellar cluster and gas disc are presented in Section~\ref{sec:method}. The effect of the AD on the inner component of the surrounding star cluster is examined in Section~\ref{sec:sc}. We scale the results to real galaxies in Section~\ref{sec:scale}, summarise and discuss the results in Section~\ref{sec:CON}.
\section{Method}
\label{sec:method}

\subsection{Simulation details}

We use an improved version of the direct $N$-body code $\phi GRAPE$ \citep{HarfstEtAl2007} including the friction force of stars in the AD. The code is parallel and uses GPU accelerators for the calculations of the gravitational force. The integration of the equation of motion is done using a 4th order Hermite scheme. For more details see Paper~I.
The $\phi GRAPE$ code was used in Paper~I and Paper~II as well as in many other papers on galactic nuclei and tidal disruption events \citep[e.g.][]{Zhong2014,Zhong2015,Li2017}.

Since the previous papers focused on the accretion rate onto the SMBH, the effects of stellar tidal disruptions were included in the simulation. It was done in a way that if an object crosses the accretion radius $r_\mathrm{acc}$ then it is considered to be tidally disrupted and 100\% of the mass is added to the mass of the SMBH. We used $r_\mathrm{acc}$ as a free numerical parameter which regulates the spatial resolution. 

Here we use the data from the most realistic simulation of Paper~II (designated as 128k03r) where the number of stars was set to $N=1.28\times10^5$, the accretion radius $r_\mathrm{acc} = 3.0\times 10^{-4}r_\mathrm{inf}$, where $r_\mathrm{inf}$ is the influence radius of the SMBH. The scale height of the AD was set to have linear dependence on radius in the inner region (see next subsection). We compare the data with the analogous 128k simulation without the AD (128k03ng). The number of particles in the simulations is still much smaller than the number of stars in a real galactic centre, so each particle represents a group of stars. A detailed description of the scaling procedure of star--disc interactions is given in Paper~I.

We use H\'enon units (also known as $N$-body units) throughout the paper unless otherwise specified. The total mass of the NSC as well as the gravitational constant $G$ are set to unity. We set the initial mass of the SMBH and the AD to be 10\%  and 1\% of the total stellar mass, respectively. The SMBH grows due to the capture of stars while the AD remains stationary. 

\subsection{Disc model}

Our model of the AD corresponds to an axisymmetric thin disc based on \citet{SS1973} and \citet{NT1973}. The gas density is given by
\begin{eqnarray}
\rho_\mathrm{g}(R,z) & = & \frac{2-p}{2\pi \sqrt{2 \pi}} \frac{M_\mathrm{d}}{h R_\mathrm{d}^3} \left( \frac{R}{R_\mathrm{d}} \right)^{-p}\nonumber\\
& & \exp \left[ -\beta_s \left( \frac{R}{R_\mathrm{d}} \right)^s \right] \exp \left( \frac{-z^2}{2 h^2} \right),
\label{gasdensity}
\end{eqnarray}
where $p=3/4$ is the surface density power-law index corresponding to the outer region of a standard thin disc model, $R$ is the radial distance from the SMBH, $z$ is the vertical distance from the disc plane, $R_\mathrm{d} = 0.22$ is the radial extent of the disc (scaled with the influence radius of the SMBH). The parameters $s=4$ and $\beta_s=0.7$ are associated with the smoothness of the outer cutoff of the disc (introduced for numerical reasons) and $h$ is the scale height. The gas in the disc is set to have a Keplerian rotation profile. The total disc mass is fixed to be $M_\mathrm{d} = 0.01$ and the gravity force from the AD is neglected.

We approximate the disc scale height in the inner region with a linear relation $h = \frac{R}{R_\mathrm{sg}}h_\mathrm{z}$ up to a distance $R_\mathrm{sg} \approx 0.026$ where the vertical self-gravity of the disc becomes important. With $h_\mathrm{z}=2.2\,10^{-4}$ the opening angle of the AD is 0.5$^{\circ}$. In the region of a vertically self-gravitating disc, the scale height is constant $h = h_\mathrm{z}$. The transition between two regions is estimated by equating the vertical component of the spherically symmetric force from the SMBH at $z=h_\mathrm{z}$ with the vertical self-gravitation of a thin disc above the AD. We examined the effects of changing the inner disc height profile on the results in Paper~II. 

\subsection{Stellar component and star--disc interactions}

The initial conditions are generated in the following way. We place a point-mass potential into a Plummer sphere (with virial radius of one H\'enon unit) and evolve the system to the stage of dynamical equilibrium $t=0.001t_\mathrm{rel}$ (several crossing times). After that, the influence radius of the SMBH (the enclosed radius where the total stellar mass equals that of the SMBH) is measured to be $r_\mathrm{inf} = 0.22$. The NSC consists of equal mass stars. Then the interaction with the AD is `switched on'. The total simulation time is 2 half mass relaxation times ($t_\mathrm{rel}$).

Since we neglect the gravity of the AD, a star feels the gas as a drag force which is given by the equation
\begin{equation}
{F}_\mathrm{drag} = - Q_\mathrm{d} \pi r_\star^2 \rho_\mathrm{g}(R,z) \left| {V}_\mathrm{rel} \right| {V}_\mathrm{rel},\label{eq:fdrag_star}
\end{equation}
where $\rho_\mathrm{g}$ is the local gas density (equation \ref{gasdensity}), $r_\star$ is the stellar radius and ${V}_\mathrm{rel}$ is the relative velocity between the star and the gas, $Q_\mathrm{d}$ is the drag coefficient, we use $Q_\mathrm{d}=5$ \citep{CourantFriedrichs1948}. We assume that stars have supersonic motion while crossing the disc and therefore we treat the drag as a ram pressure effect. The contribution from dynamical friction is neglected since it is proportional to $V_\mathrm{rel}^{-2}$ while the ram pressure drag goes with $V_\mathrm{rel}^2$ (see \citealt{Ostriker1999} and Sec. 2.2 of Paper~I).

Since a star particle represents a group of stars, we scale equation~(\ref{eq:fdrag_star}) by introducing an effective dissipative parameter,
\begin{equation}
Q_\mathrm{tot}(N) \equiv Q_\mathrm{d} N \left( \frac{r_\star}{R_\mathrm{d}}\right)^2.\label{Qtotdef}
\end{equation}
This expression describes the dimensionless total dissipative cross section of $N$ stars, normalised to the disc area. Now, Eq. \ref{eq:fdrag_star} can be rewritten as acceleration in terms of global quantities, such as $R_\mathrm{d}$, $M_\mathrm{cl}$ and $Q_\mathrm{tot}$ by
\begin{equation}
a_\mathrm{d} = -Q_\mathrm{tot}\frac{\pi R_\mathrm{d}^2 \rho_\mathrm{g}}{M_\mathrm{cl}}\left| {V}_\mathrm{rel} \right| {V}_\mathrm{rel},\label{eq:adrag}
\end{equation}
where $M_\mathrm{cl}$ is the total stellar mass. To get around the fact that the relaxation time in the modelled system is shorter than the $t_\mathrm{rel}$ for a real galactic nucleus, we choose $Q_\mathrm{tot}$ in such a way that the ratio between the dissipation time-scale and the relaxation time is conserved. Thus, given a galactic centre with $N_\mathrm{real}$ stars and an effective dissipative parameter $Q_\mathrm{tot}(N_\mathrm{real})$, the value of $Q_\mathrm{tot}(N_\mathrm{sim})$ to be used in a simulation with $N_\mathrm{sim}$ super-particles is
\begin{equation}
Q_\mathrm{tot}(N_\mathrm{sim}) = \frac{t_\mathrm{rel}(N_\mathrm{real})}{t_\mathrm{rel}(N_\mathrm{sim})} Q_\mathrm{tot}(N_\mathrm{real}).
\label{QtotNrel}
\end{equation}
In our simulations the scaling for arbitrary $N$ is given by (Eq. 11 of Paper~II)
\begin{equation}
\label{Qtot_scale}
Q_\mathrm{tot}(N) \approx 5.42\ln(0.4N)/N
\end{equation}
leading to
\begin{equation}
\label{eta_def}
\eta = t_\mathrm{diss}/t_\mathrm{rel} \approx 10
\end{equation}
(see Eq. 16 and Fig. 6 of Paper I). 

For a given galaxy and AD properties (mass and size) the appropriate value of $Q_\mathrm{tot}$ can be calculated (see Table 1 of Paper~I). The $Q_\mathrm{tot}$ value of our simulation (Eq. \ref{Qtot_scale}) is consistent with the down-scaled value for M87. We refer to Paper~I for a detailed description. 

The effects from stellar crossings, gravity force from the AD mass and the contribution of stellar winds on the AD are neglected. Some of these assumptions are discussed in Sec.~\ref{sec:CON}, but a detailed analysis will be done in future work.

\section{Effect on the surrounding star cluster}
\label{sec:sc}

In our previous work (Paper~II), we analysed the statistics of orbital parameters of accreted particles and found that the paths that they take to accretion depend on their final eccentricities and inclinations. We identified three broad paths or \textit{plunge types}, these are (1) disc capture, (2) gas assisted accretion, and (3) direct accretion. The plunge type 1 stars were captured by the AD and went through a disc migration phase. 
   
\begin{figure}
\begin{centering}
\includegraphics[width=\columnwidth]{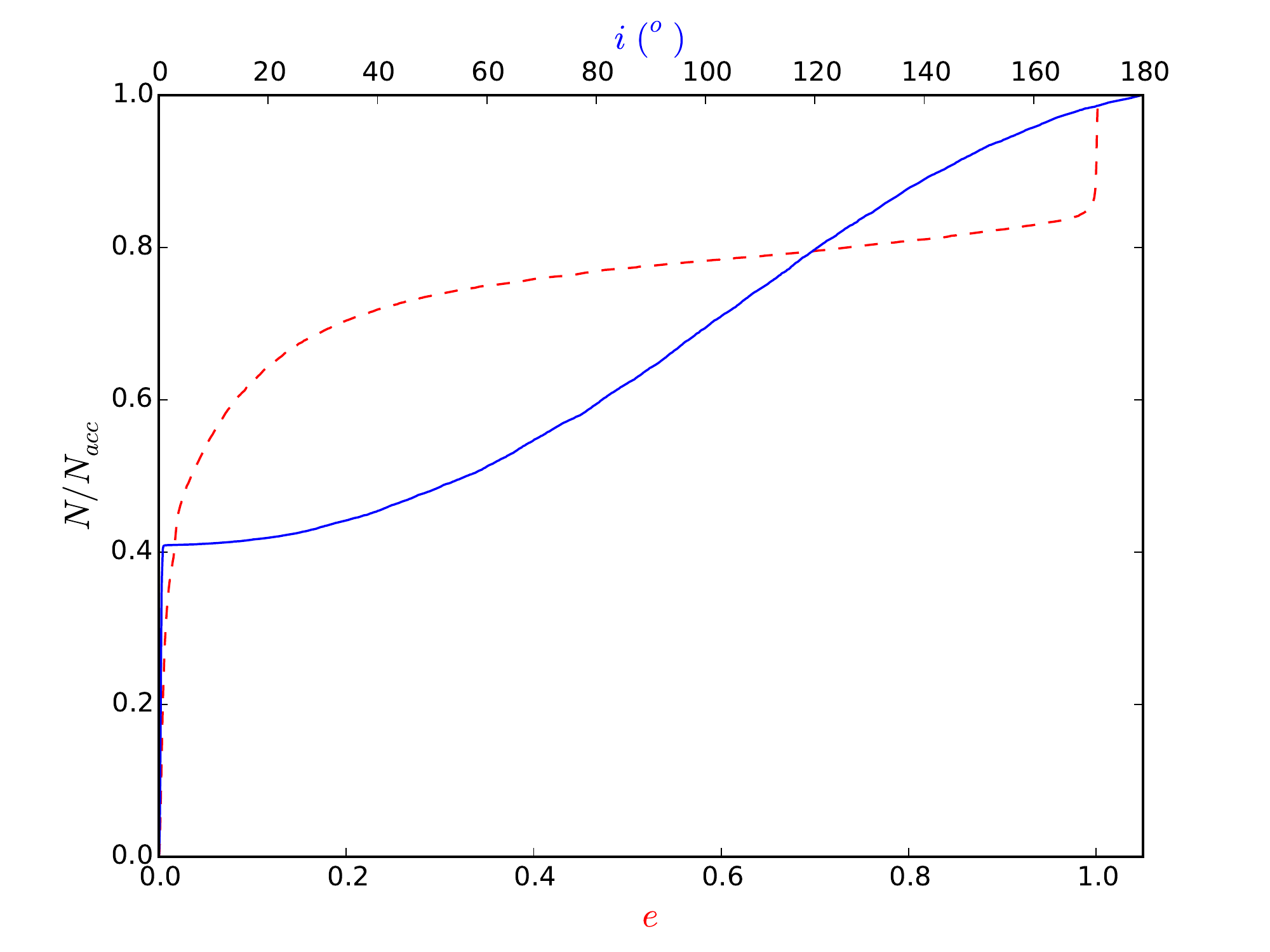} \\ 
\par\end{centering}
\caption{Cumulative distribution of the orbital parameters of all accreted particles at the moment of accretion. The red line represents the eccentricity distribution and corresponds to the bottom X-axis. The blue line represents the inclination angle distribution and corresponds to the top X-axis.}
\label{fig:EccInc}

\end{figure}

Figure \ref{fig:EccInc} shows the cumulative distribution of inclination $i$ and eccentricity $e$ at the time of accretion. The blue and red lines in Fig. \ref{fig:EccInc} clearly show that about 40\% of all accreted particles throughout the simulation are accreted with very low inclinations and eccentricities meaning the accretion through the AD (plunge type 1). As we will show, these particles were accreted only after several orbital times of residency inside the disc. While in the migration phase, the particles form a NSD as we show later. 
Moreover, the stellar disc remains stationary during the simulation and is supported by a constant inflow of stars from the outer parts of the NSC. In the following section, we examine the properties of the NSD and the stellar migration time-scale. 

\subsection{Spatial distribution of the NSD particles}

\begin{figure}
\begin{centering}
\includegraphics[width=\columnwidth]{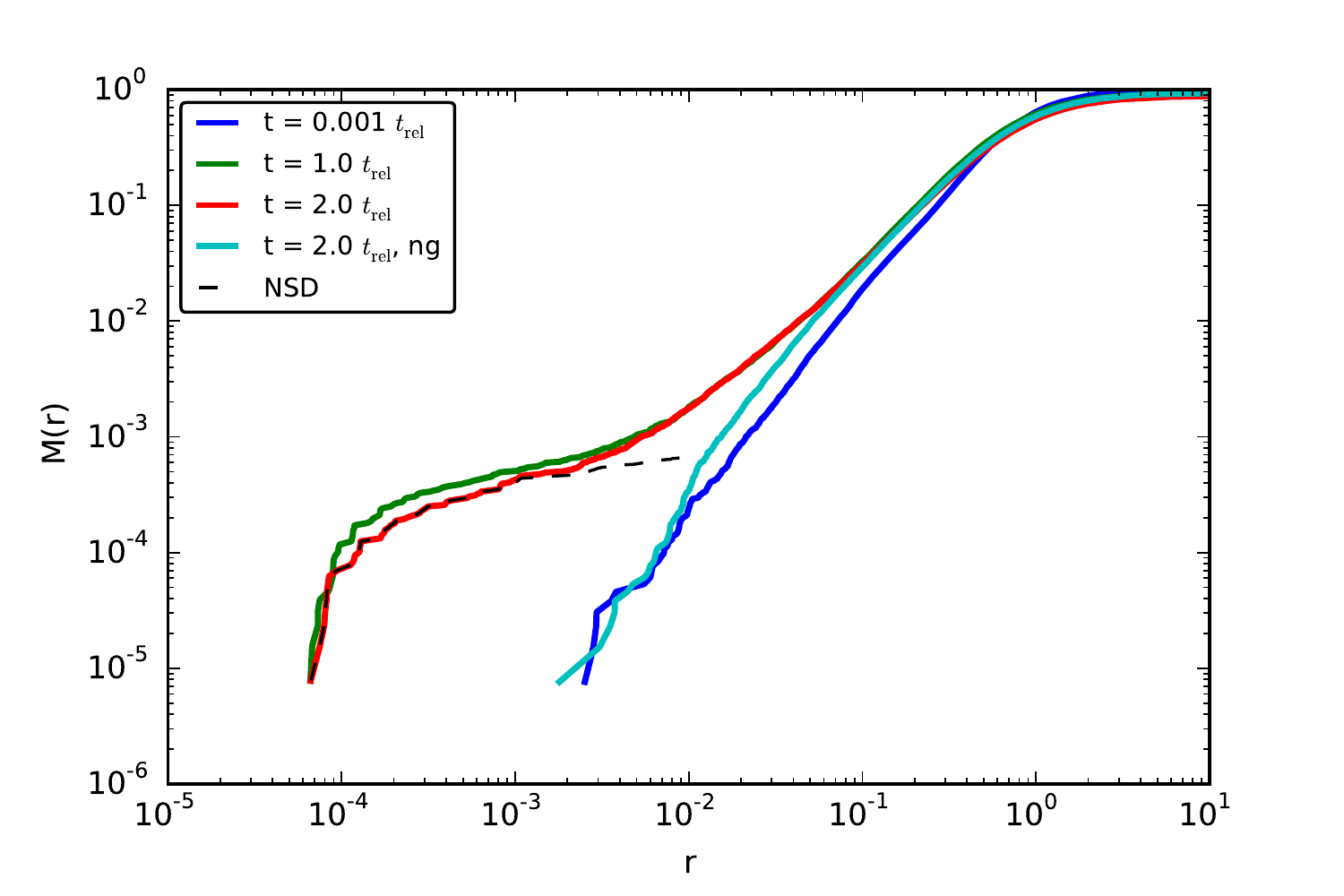} \\ 
\par\end{centering}
\caption{Evolution of the NSC in terms of the cumulative mass profiles. Blue and cyan lines represent the initial model and the model without the AD at the end of the simulation, red and green lines show the NSC mass profiles at 1 and 2 $t_\mathrm{rel}$ respectively. The black dashed line is the cumulative mass of the stellar disc.}
\label{fig:MassProf}

\end{figure}

The initial mass profile as well as the mass--radius dependence for the models with and without the AD at $t=t_\mathrm{rel}$ are presented in Fig. \ref{fig:MassProf}. The cumulative mass profiles of the NSC at 1  and 2 relaxation times reveal the mass concentration in the inner part of the cluster, the profile remains very similar until 2 relaxation times.  Although the stellar accretion from disc-captured particles occurs, we see no change in the mass profile after the NSD has formed. That means that the NSD is continuously supplied from the NSC. The mass of the NSD stays the same in order of magnitude and equals to $M_\mathrm{NSD} \approx 7.0\times 10^{-4}$ at the end of the simulation.

\begin{figure}
	\begin{centering}$\begin{array}{c}
		\multicolumn{1}{l}{\mbox{(a) $t = 0.001 t_\mathrm{rel}$}}\\
		\includegraphics[width=0.8\columnwidth]{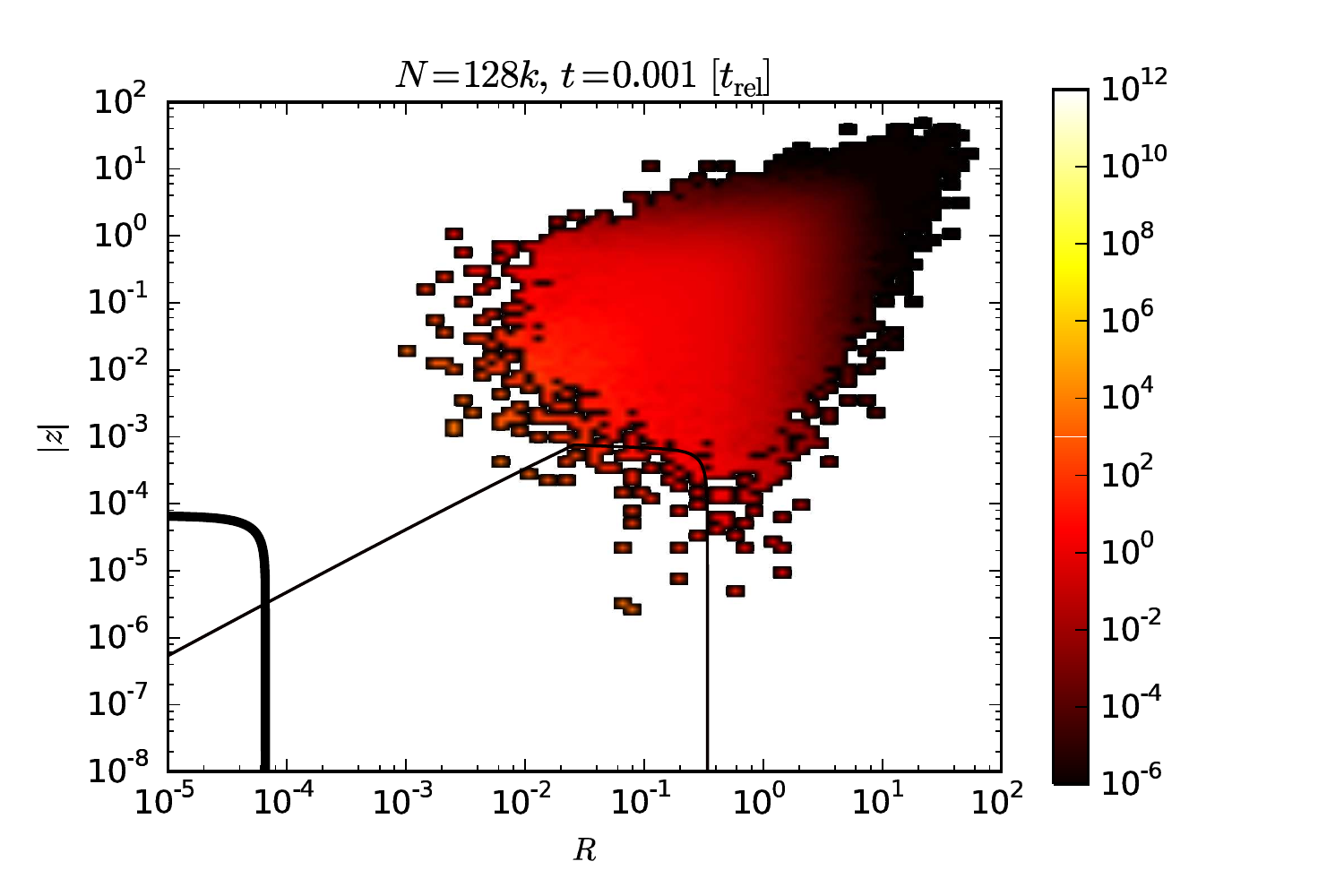}\\
		\multicolumn{1}{c}{\mbox{}}\\
		\multicolumn{1}{l}{\mbox{(b) $ t = 2 t_\mathrm{rel}$, without AD }}\\
		\includegraphics[width=0.8\columnwidth]{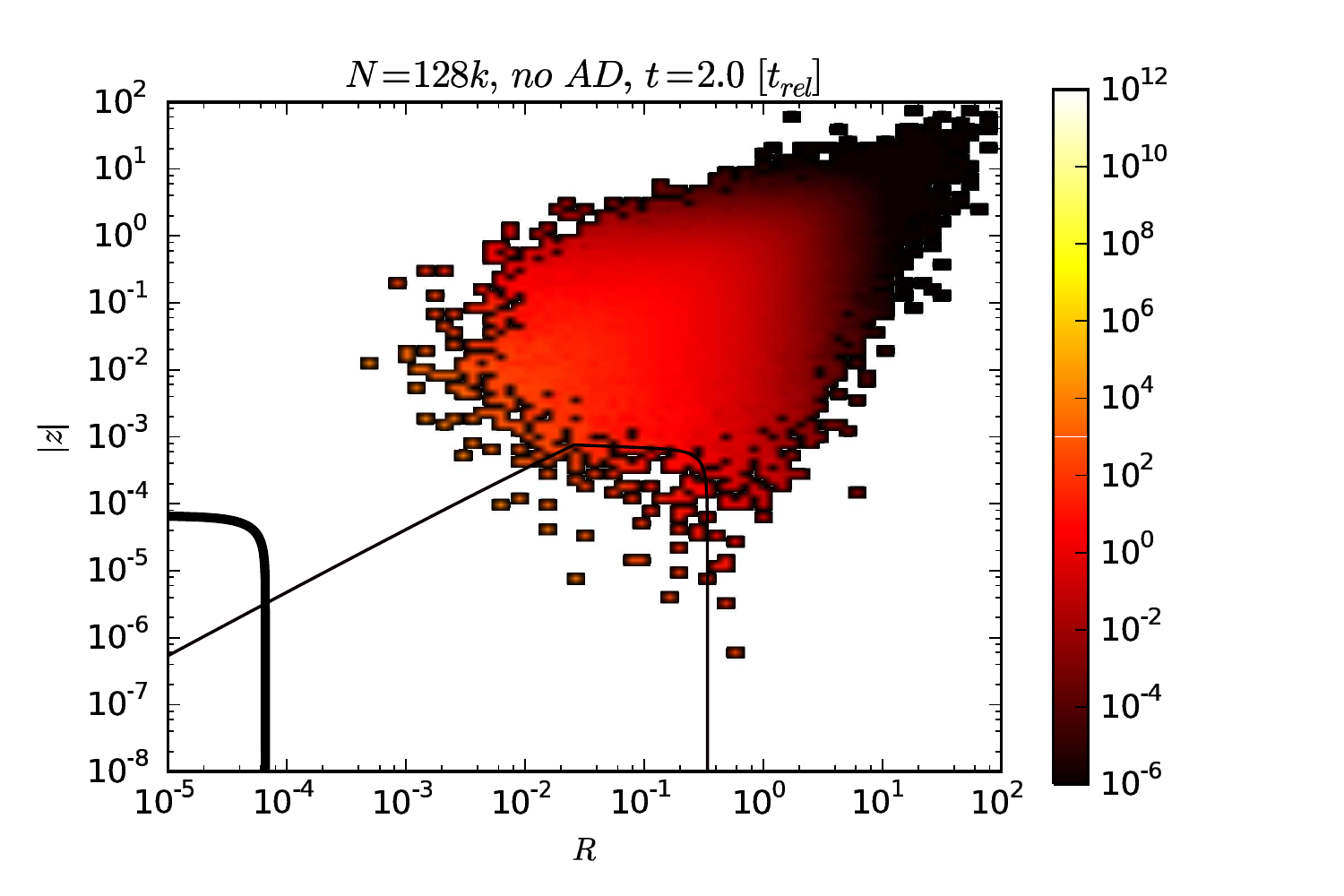}\\
		\multicolumn{1}{l}{\mbox{(c) $ t = 2 t_\mathrm{rel}$, with AD }}\\
		\includegraphics[width=0.8\columnwidth]{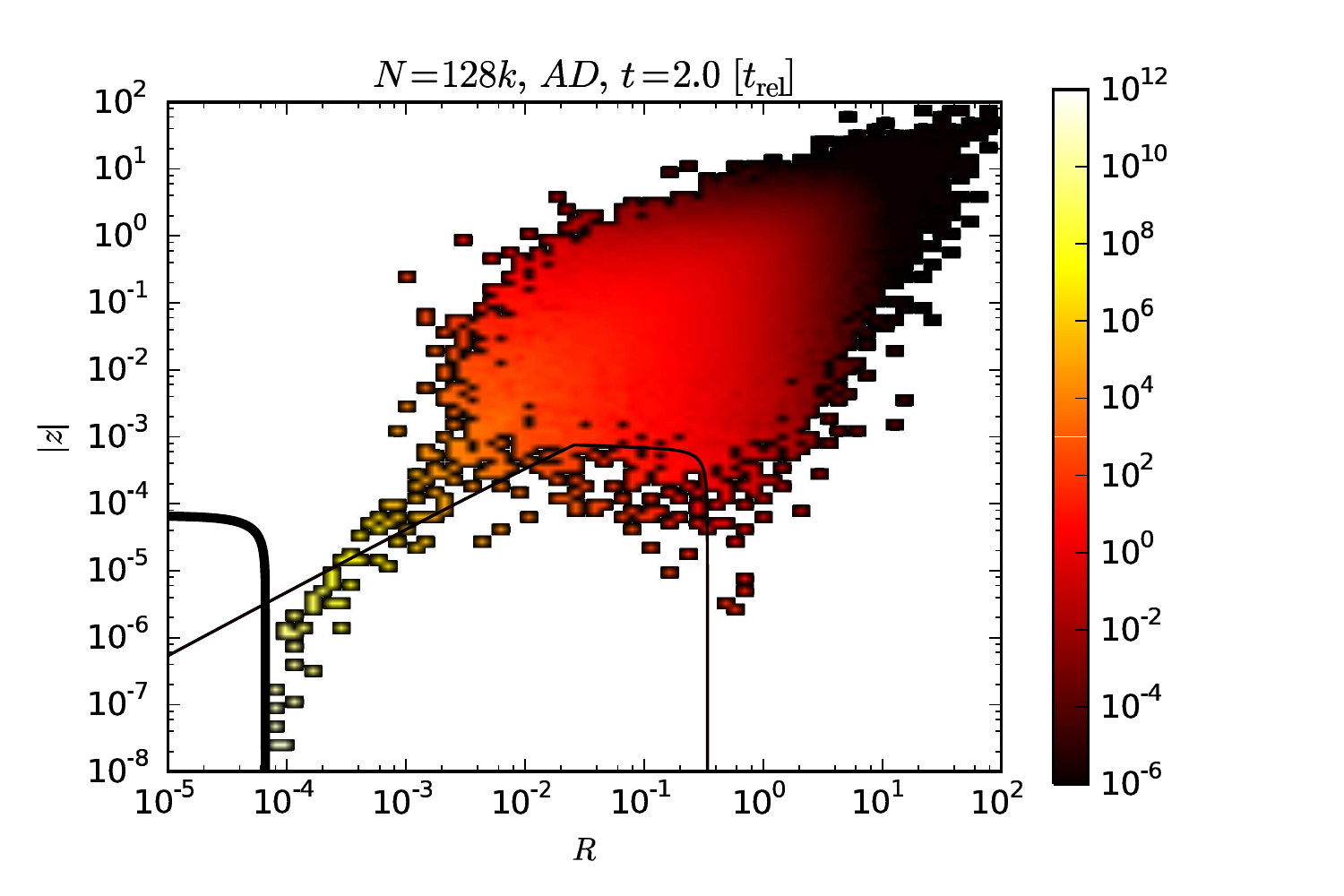}\\
		\end{array}$
		\par\end{centering}
	\caption{Spatial density distribution of the NSC. Panel (a) is the initial model after 5 crossing times. The other two panels correspond to the end of simulation. Panel (b) shows the evolution of the model without AD. The bottom panel (c) shows the formed disc-like substructure of the NSC due to the AD. The thick black line represents the accretion radius $r_\mathrm{acc}$ and the thin line stands for the AD density $\rho=1$ indicating the boundary of the gas disc. The colour-code indicates the stellar volume density.}
	\label{fig:SCDens}
\end{figure}

\begin{figure}
\begin{centering}
\includegraphics[width=\columnwidth]{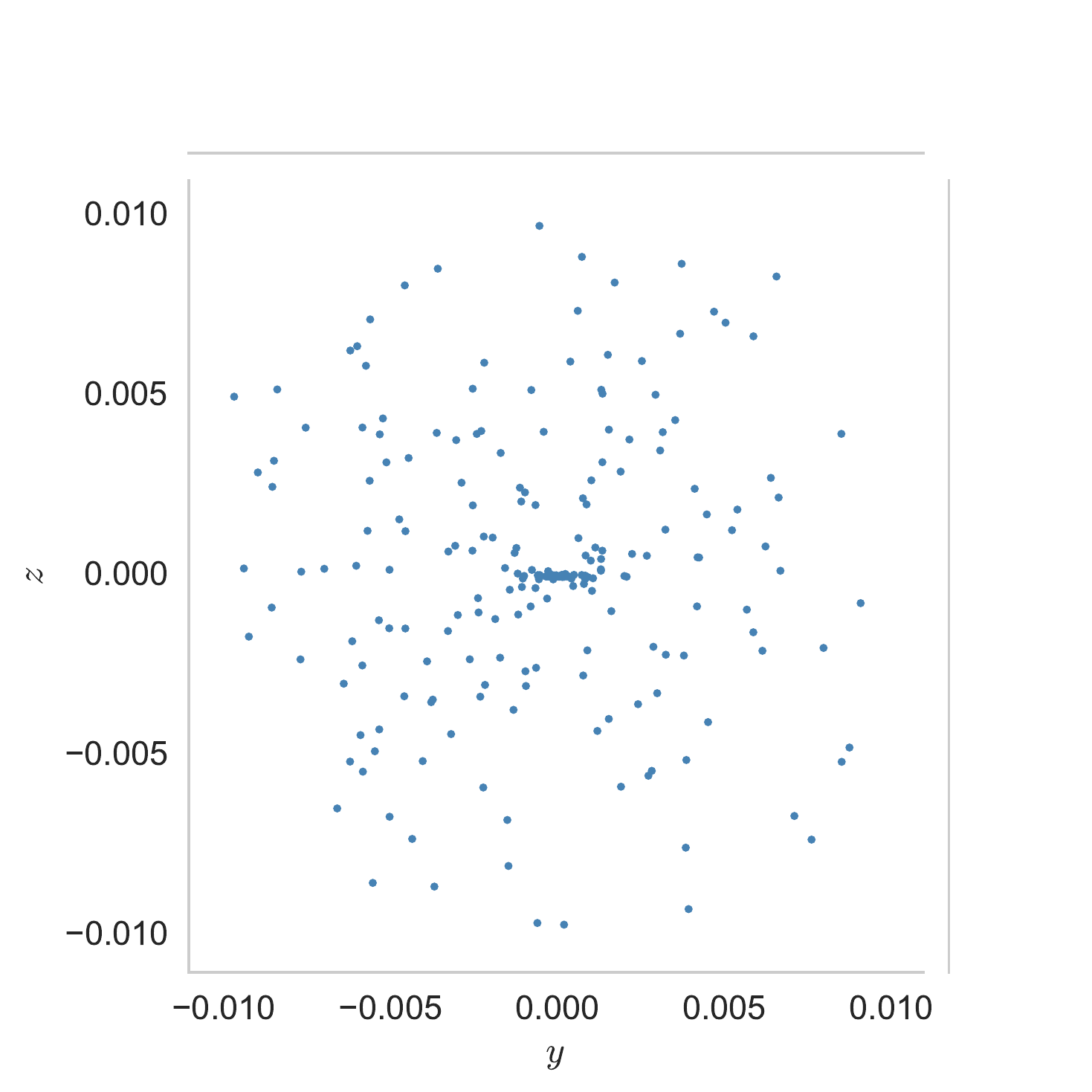}\\
\par\end{centering}
\caption{Spatial distribution of stars inside $r = 0.05\times R_\mathrm{d}$ in the YZ plane at time $t = 1.0t_\mathrm{rel}$. At the centre a disc-like structure has formed due to star--disc interactions.}
\label{fig:NSCdisc}
\end{figure}

Figure \ref{fig:SCDens} shows the spatial density distribution of the NSC. As we can see from the Fig. \ref{fig:SCDens} the initial Plummer sphere (looks like a triangle in the logarithmic $z$ vs $R$ plane) is slightly distorted inwards due to the presence of the SMBH, while the AD leads to the formation of a `tail' of stars in the innermost part of the cluster with $|z| \ll R$. The `tail' can be clearly visualised as a disc of stars (Fig. \ref{fig:NSCdisc} ). We call the disc of stars the \textit{nuclear stellar disc} (NSD). 

In order to investigate the properties of the NSD particles and the NSD as a whole, we have to define the criterion for a stellar disc particle. The `tail' in Fig. \ref{fig:SCDens} panel (c) gives us constrains on the vertical and radial distances of the NSD particles. Comparing panels (b) and (c) of the same figure we set the condition: $R<10^{-2}$ $\&$ $z < 10^{-3} $. Furthermore we require $e < 0.5$ in order to remove transient particles and since the orbital inclination angle is derived as $\cos i = \frac{L_z}{|L|}$, the condition $\cos i > 0.0$ excludes the counter-rotating stars from the NSD.
Putting all together we define the NSD particles as particles that satisfy the following criteria:   
\begin{eqnarray}
\label{disc_cond}
R  <   10^{-2};\nonumber
|z|  <  10^{-3};\nonumber\\
\cos(i)  >  0.0;
e  <  0.5.
\end{eqnarray}
$R,z$ are cylindrical coordinates. 

\begin{figure}
	\begin{centering}
		\includegraphics[width=\columnwidth]{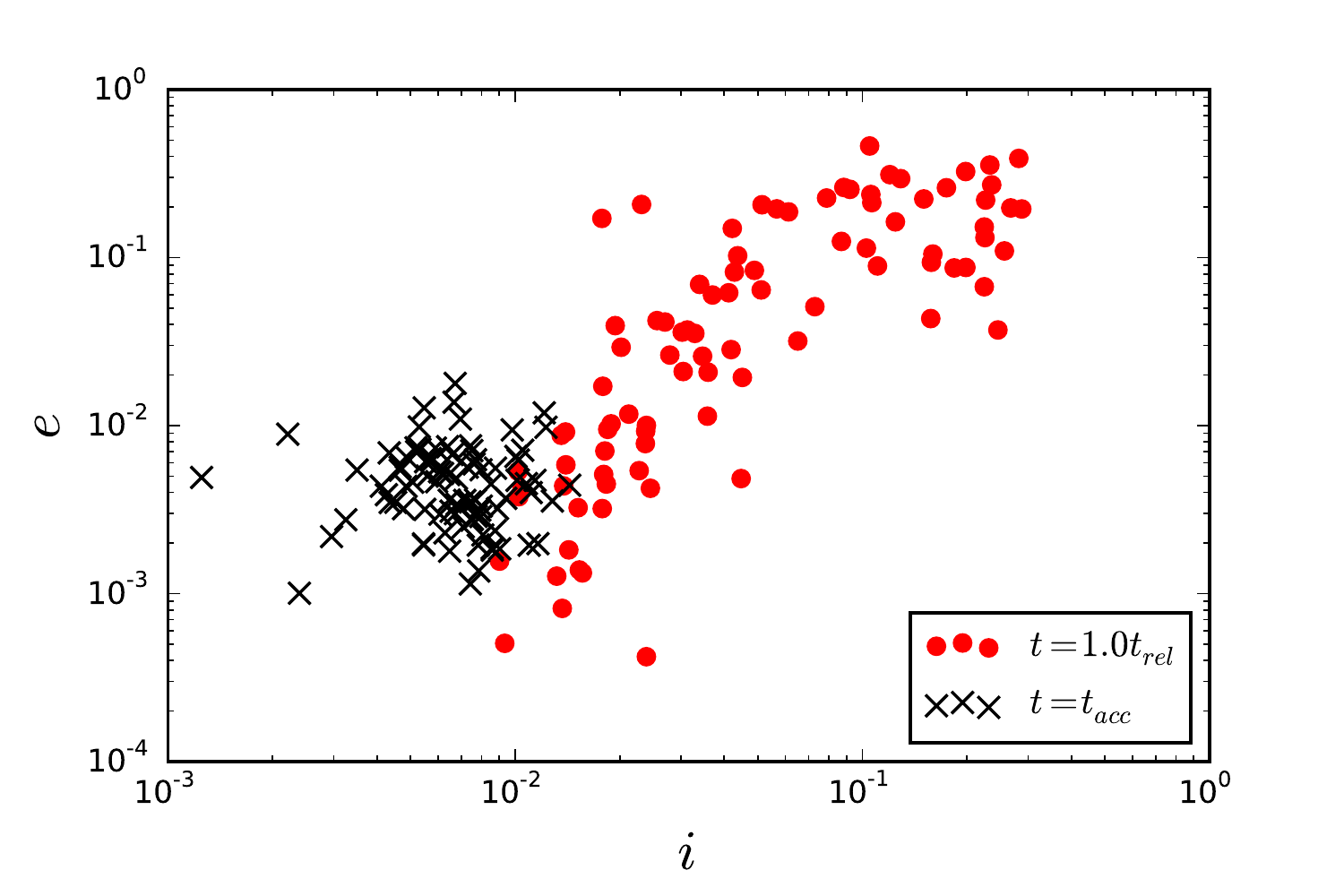}\\
		\par\end{centering}
	\caption{Distribution of the NSD particles in the eccentricity--inclination plane at $t_\mathrm{cur} = 1.0 t_\mathrm{rel}$ (red circles). The black crosses are the eccentricity and inclination values of those stars at their time of accretion.}
	\label{fig:eidisc}
\end{figure}

We ensure that these criteria select plunge type 1 stars by plotting the distribution of selected NSD particles in the eccentricity--inclination plane when they still live in the stellar disc ($t = 1.0 t_\mathrm{rel}$) and when they were accreted onto the SMBH ($t = t_\mathrm{acc}$). Fig. \ref{fig:eidisc} shows that all these particles were accreted with very low values of eccentricity and orbital inclination.
The size of the NSD is $\approx$ 3 times smaller than the effective radius of the AD ($R_\mathrm{eff}=0.032$ in Paper~II; a characteristic location in the AD where most stars begin their plunge).

\begin{figure}
\begin{centering}
\includegraphics[width=\columnwidth]{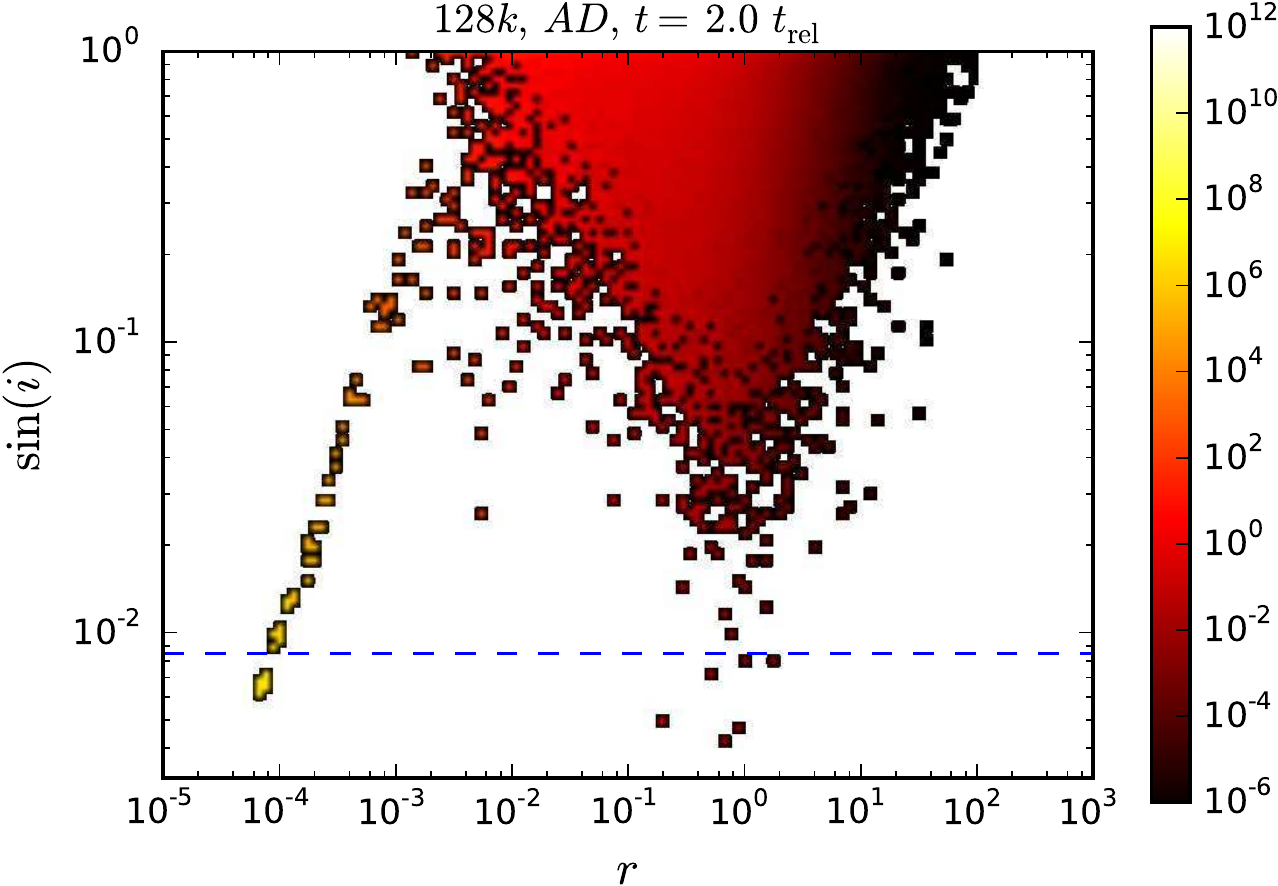}\\
\par\end{centering}
\caption{Inclination angles of all stars as function of distance $r$ from the SMBH coloured by stellar density. The blue dashed line indicates the opening angle of the AD.}
\label{fig:sin}
\end{figure}

\begin{figure}
	\begin{centering}
		\includegraphics[width=\columnwidth]{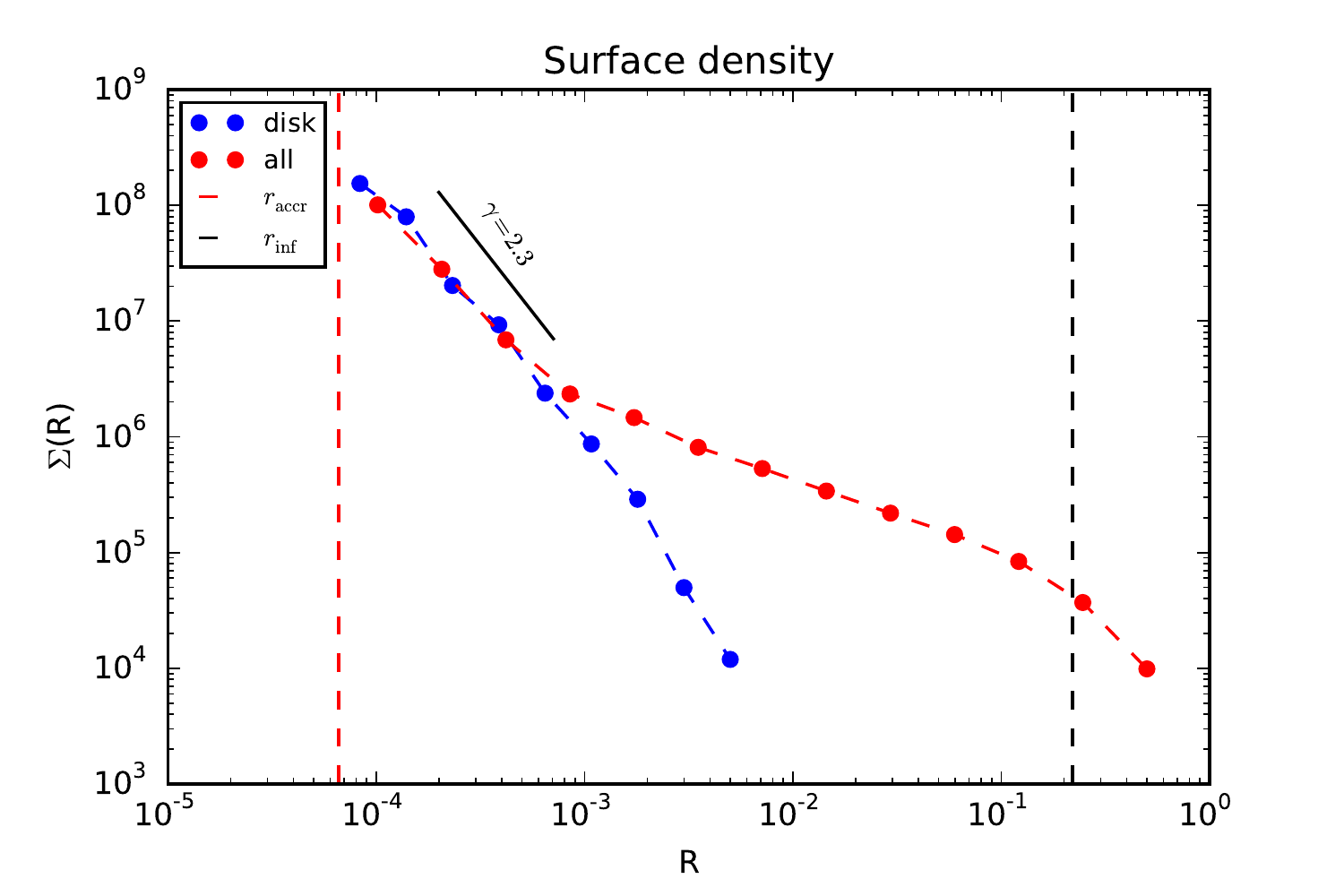}\\
		\par\end{centering}
	\caption{Surface density of the NSD. The red dotted line represents all NSC stars, the blue one shows only the stars that belong to NSD. Dashed red and black vertical lines represent the accretion radius and the influence radius respectively. }
	\label{fig:surfden}
\end{figure}

Figure \ref{fig:sin} shows that the inclination declines approximately proportional to the size of the orbit as expected due to the friction force in a Kepler rotating AD.
The surface number density of the NSD at $t = t_\mathrm{rel}$ is displayed in Fig. \ref{fig:surfden}. The figure shows a strong overdensity in the inner region ($r<10^{-3}$) of the NSC. The surface density features a steep power law profile with $\gamma=2.3$.

\subsection{Lifetime and evolution of the NSD}

\begin{figure}
	\begin{centering}
		\includegraphics[width=\columnwidth]{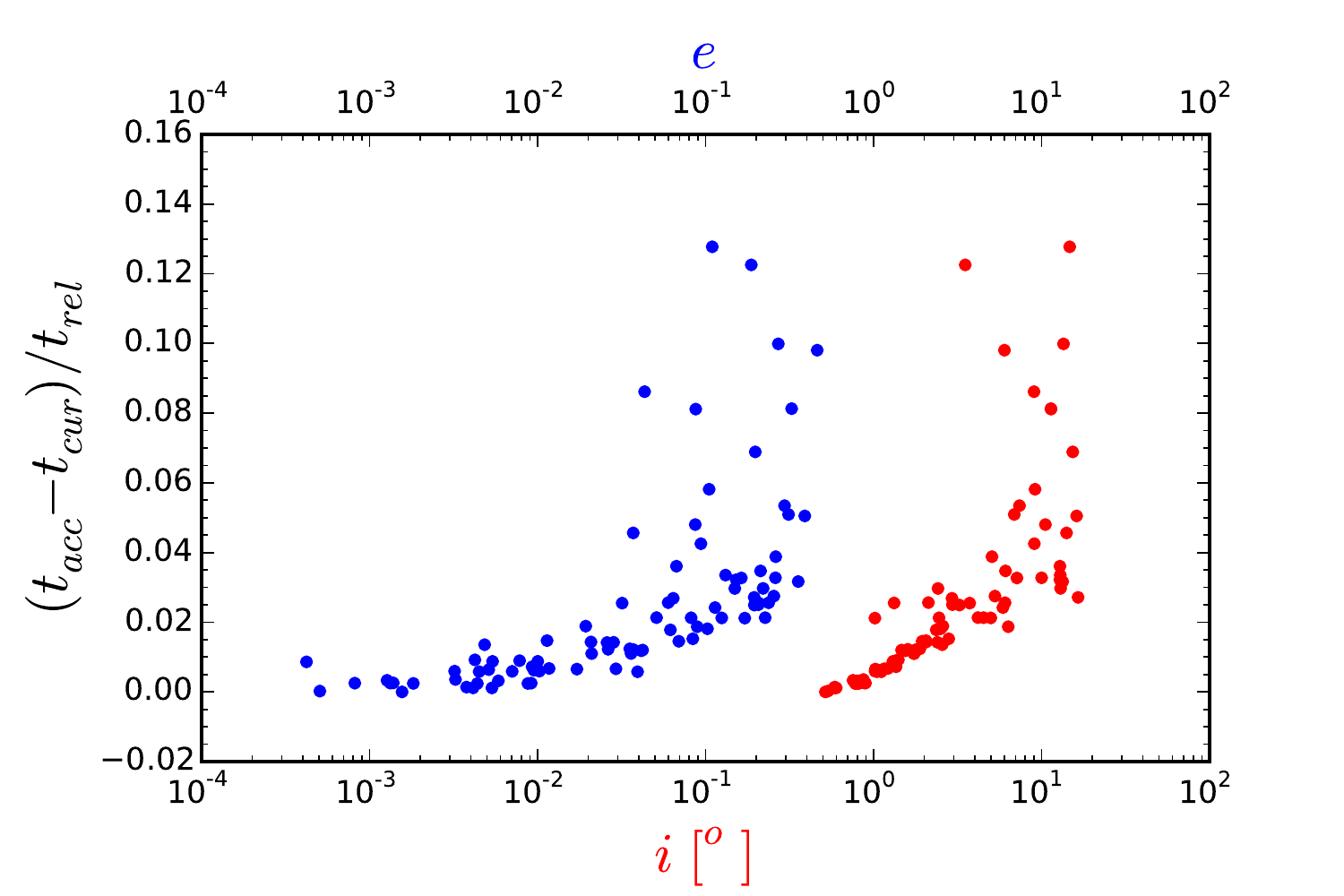}\\
		\par\end{centering}
	\caption{The time intervals left to accretion as function of inclination (lower horizontal axis) and eccentricity (upper horizontal axis) at $t = 1.0\,t_\mathrm{rel}$ of the disc particles.}
	\label{fig:eidisc_life}
\end{figure}

\begin{figure}
	\begin{centering}
		\includegraphics[width=\columnwidth]{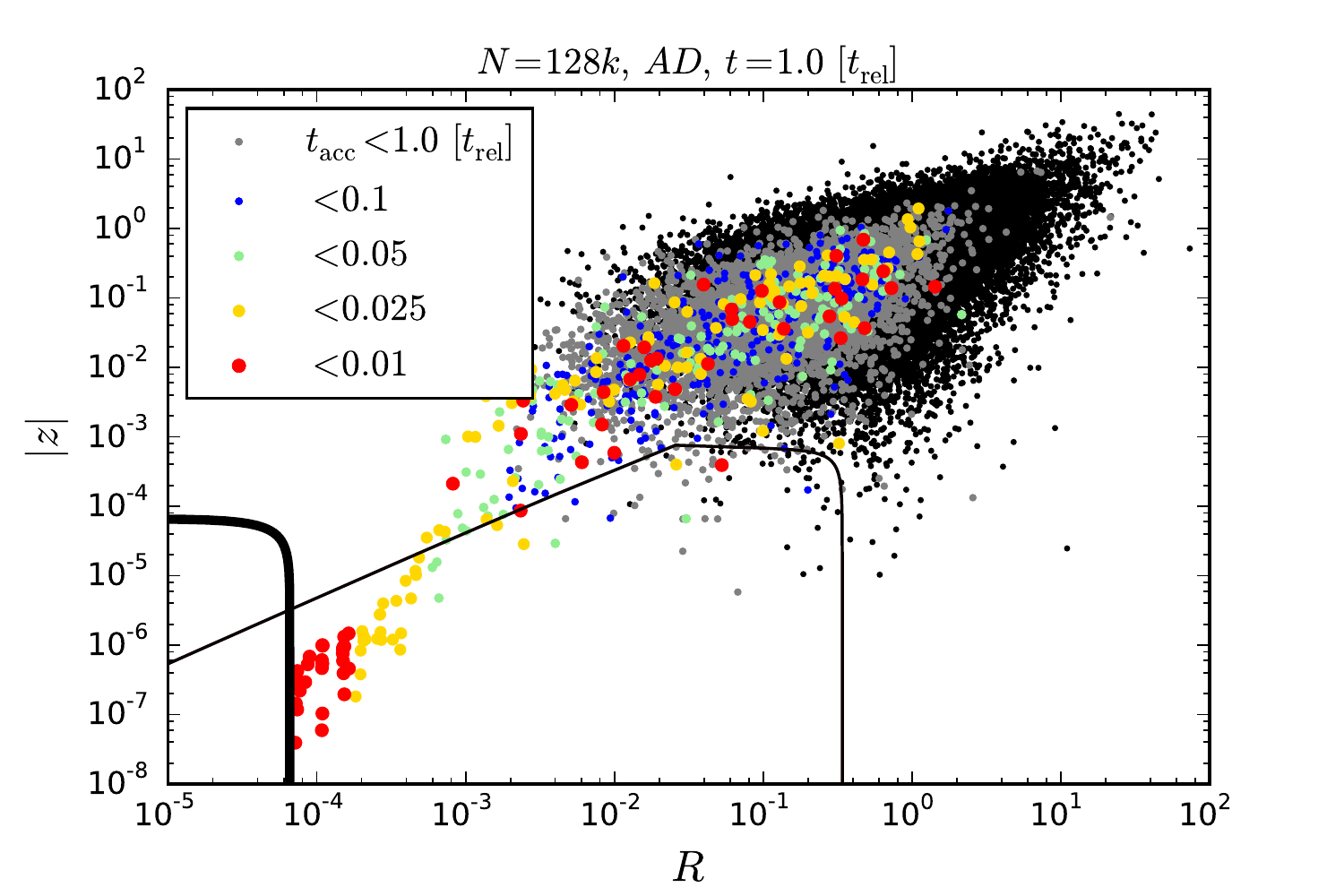}\\
		\par\end{centering}
	\caption{Spatial distribution of all stars in the NSC at $t=1.0\,t_\mathrm{rel}$ coloured by the time left until accretion. Thick and thin black lines are the same as in Fig. \ref{fig:SCDens}.}
	\label{fig:timeline}
\end{figure}

\begin{figure}
	\begin{centering}
		\includegraphics[width=\columnwidth]{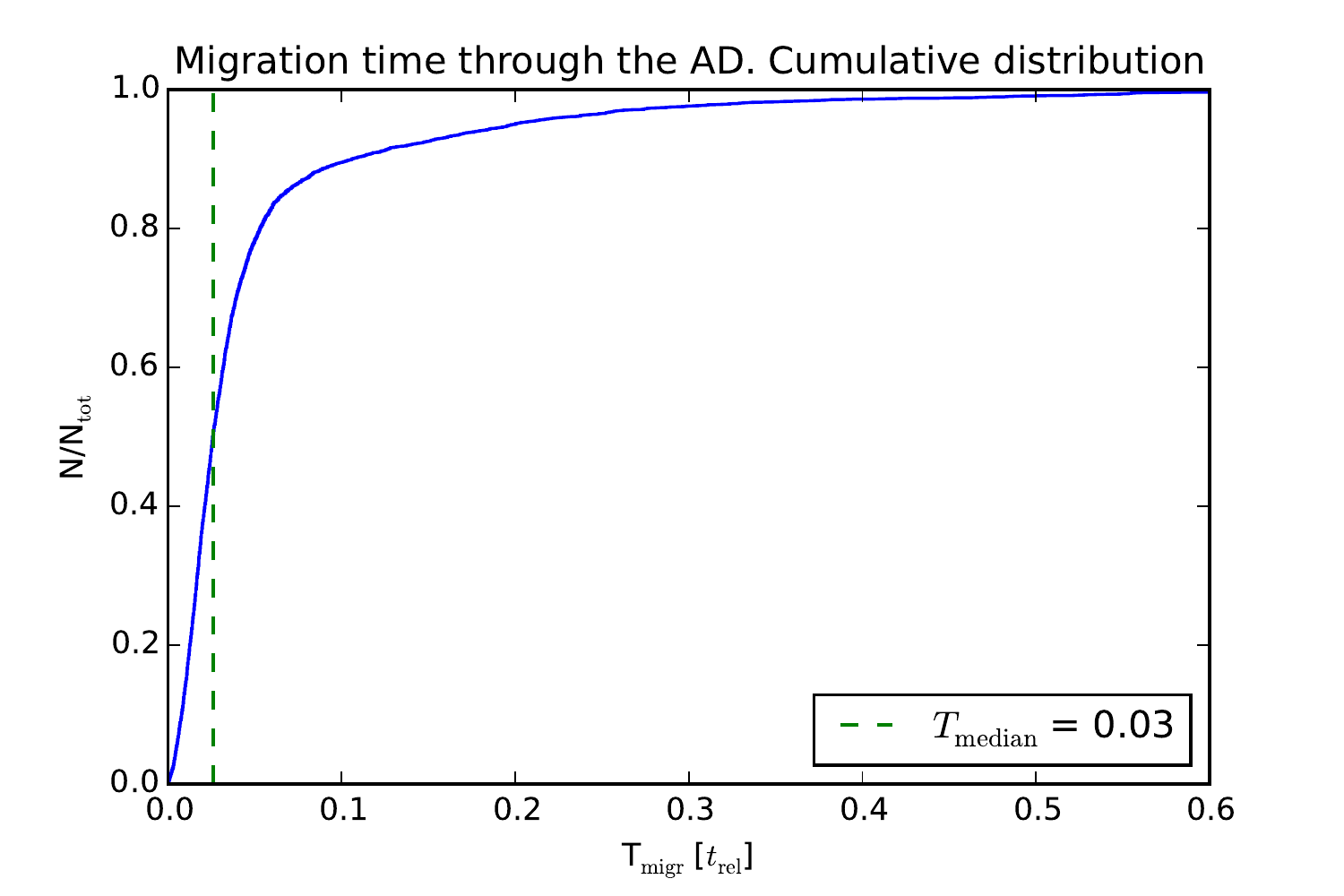}\\
		\par\end{centering}
	\caption{Cumulative histogram of the time intervals from the moment of capture by the AD until the accretion to the SMBH. Dashed vertical lines represent median and mean time correspondingly.}
	\label{fig:tmigr}
\end{figure}

A look at some properties of the NSD stars at some arbitrary current time $t_\mathrm{cur}$ and the time left until accretion gives details on how and how fast these properties change during the migration phase. 
Figure \ref{fig:eidisc_life} shows the time left for accretion against the orbital parameters of the NSD particles at time $t_\mathrm{cur}=1.0\,t_\mathrm{rel}$ and demonstrates the decay of eccentricities and inclinations in the migration phase. All the NSD particles accrete in a fraction of the relaxation time.
Figure \ref{fig:timeline} shows the same distribution as the bottom panel of Fig. \ref{fig:SCDens}, but colour coded by the time left to accretion for all stars in the NSC. It is clearly seen that there is an inflow of particles towards the SMBH at $R < 10^{-3}$. \\
In order to compute the time a star spends inside the AD during the plunge, we track it from the moment it is captured (equation \ref{disc_cond} is fulfilled) and follow the stellar orbit until it is gone inside the accretion radius (Fig. \ref{fig:tmigr}). We tracked all NSD particles from the beginning of the simulation up to $t=1.8t_\mathrm{rel}$ and found that the median time (between capture and accretion) equals $t_\mathrm{migr} = (0.026 \pm 0.002 )t_\mathrm{rel}$. It represents the fact that 50\% of all NSD stars spend in the disc no more time than $t_\mathrm{migr}$. On the other hand it represents the renewal time of the NSD as we show below.

\begin{figure}
	\begin{centering}
		\includegraphics[width=\columnwidth]{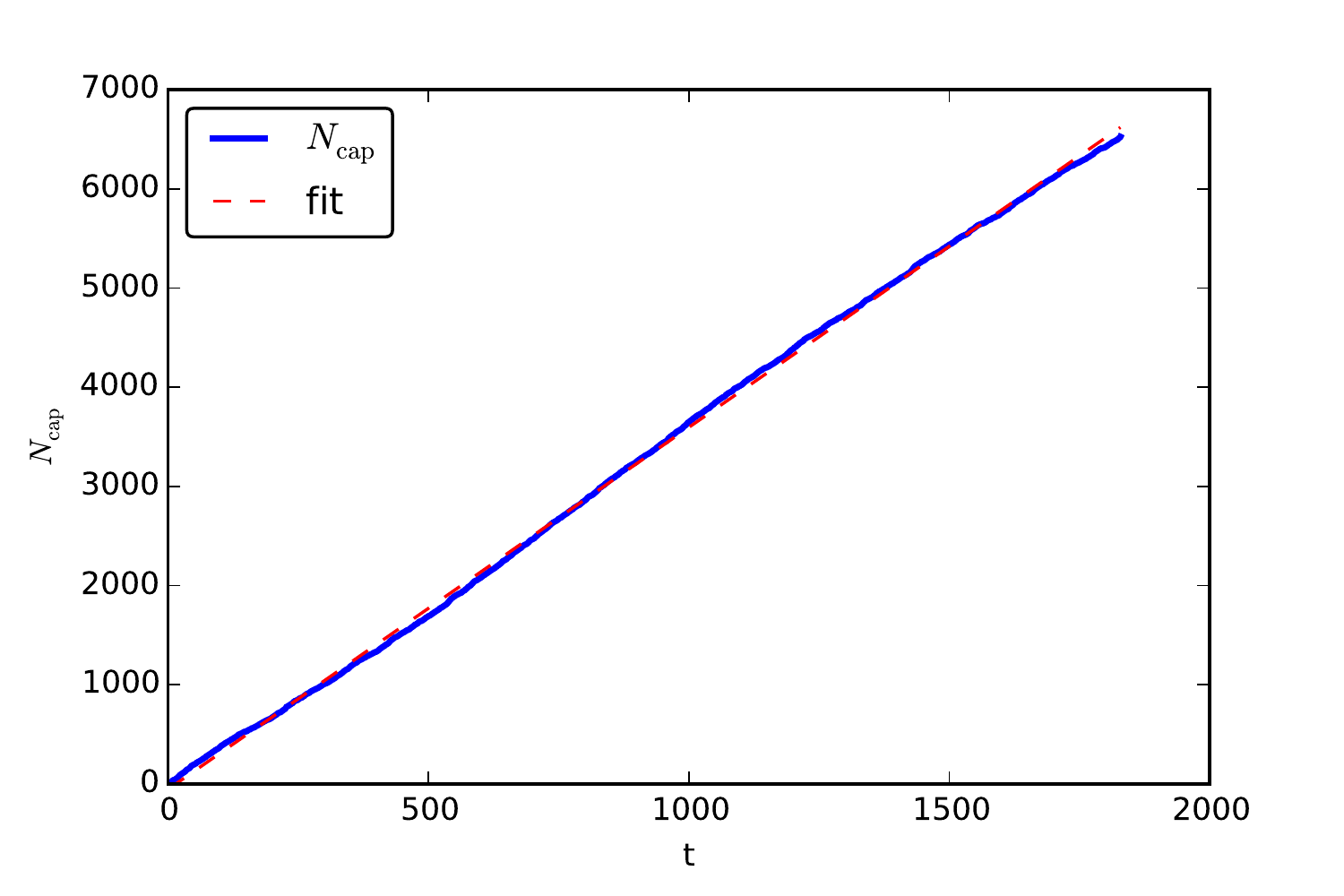}\\
		\par\end{centering}
	\caption{Cumulative number of newly captured particles by the AD. The dashed line shows a linear fit $N_\mathrm{cap} = kt$ with $k = 3.65$. }
	\label{fig:ncap}
\end{figure}

Indeed, the cumulative number of stars captured by the AD is a linear function of time (see Fig. \ref{fig:ncap}) and its derivative represents the total stellar influx
\begin{equation}
f_\mathrm{NSD} \equiv \frac{dN_\mathrm{cap}}{dt} = 3.649 \pm 0.001
\end{equation}
where $N_\mathrm{cap}$ is the cumulative number of stars captured by the NSD (but not necessarily the number of stars actually residing in the NSD, as stars eventually are accreted onto the SMBH) and $f_\mathrm{NSD}$ is the stellar flux, or the capture rate by the AD which equals $\simeq 3.65$ particles per $N$-body time unit (this is the straight line fit in Fig. \ref{fig:ncap}).

The number of stars resident in the NSD is roughly constant in time and its average value is $\langle N_\mathrm{NSD} \rangle = 81.5$ with standard deviation $8.7$. From this number and the influx rate, we can calculate a characteristic time-scale in which the entire resident stellar population of the NSD is replaced. We define this time-scale as the renewal time, which is then
\begin{equation}
t_\mathrm{renew} = \frac{\langle N_\mathrm{NSD} \rangle }{f_\mathrm{NSD}} = 22 \pm 2.
\end{equation}
As expected, $t_\mathrm{renew}$ is $\simeq 2.5\%$ of the half-mass relaxation time, and equals the migration time $t_\mathrm{migr}$ calculated previously. This time-scale is also associated with the formation time of the NSD.

\begin{figure}
	\begin{centering}
		\includegraphics[width=\columnwidth]{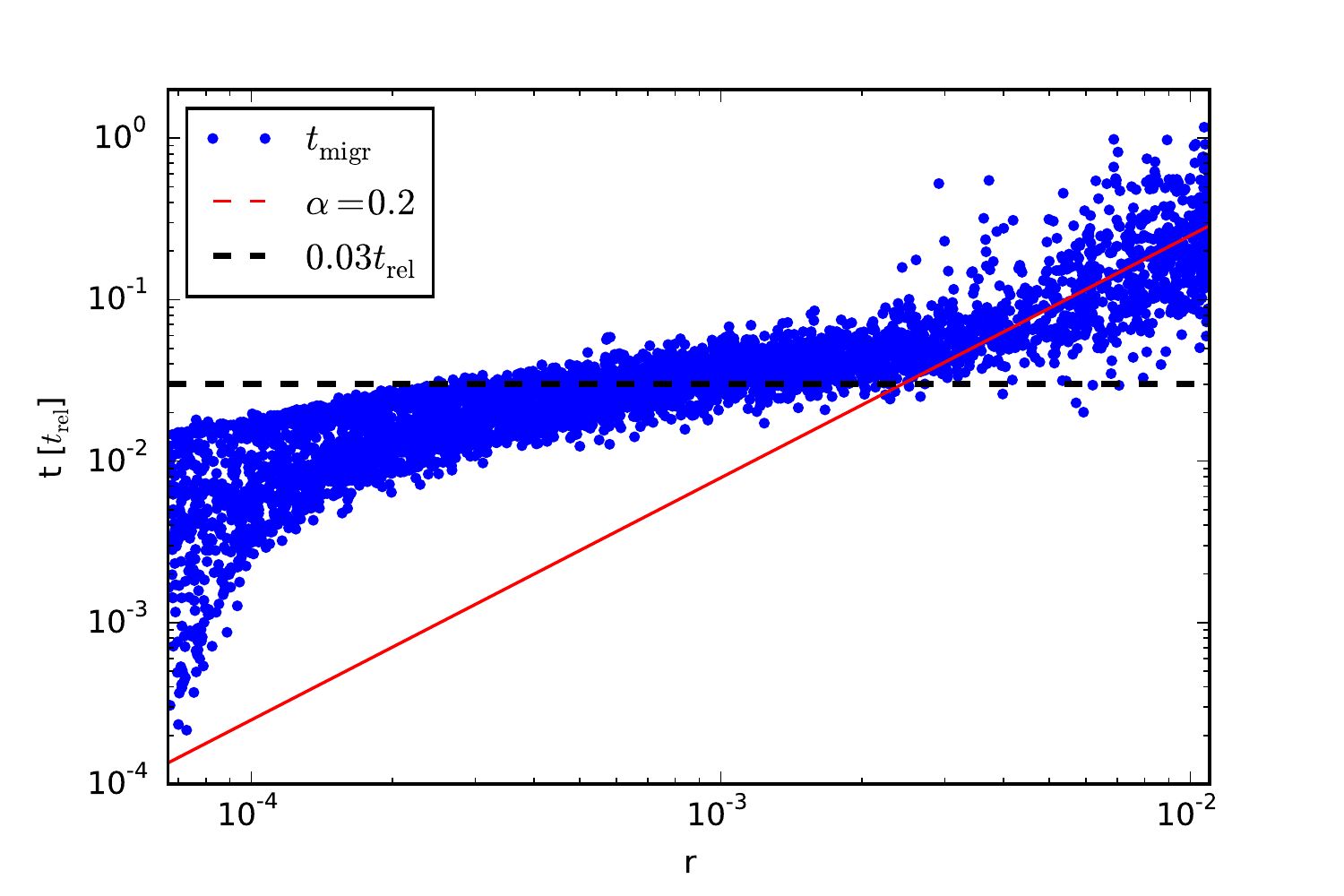}\\
		\par\end{centering}
	\caption{Migration time as a function of distance from the SMBH at which the stars `enter' the NSD. The red line represents the AD viscous time-scale with $\alpha=0.2$, the dashed line shows the effective migration time.}
	\label{fig:visc}
\end{figure}

A comparison of the migration time with the stellar dissipation time $t_\mathrm{diss} = E_\mathrm{kin}/\dot{E}_\mathrm{sd}$ (see Eq.(16) of Paper~I), where $E_\mathrm{kin}$ is the kinetic energy of all stars and $\dot{E_\mathrm{sd}}$ is the total energy dissipation rate due to the AD, shows that $t _\mathrm{diss}$ exceeds $t_\mathrm{migr}$ by two orders of magnitude. On the other hand, the viscous time-scale of the AD
\begin{equation}
\tau = \left(\frac{h}{R_\mathrm{d}}\right)^{-2}\frac{1}{\alpha\Omega}
\end{equation}
can be shorter or longer than the effective migration time and depends on the viscosity parameter $\alpha$ (Fig. \ref{fig:visc}). Here $\Omega$ is the Keplerian orbital frequency. 

\begin{figure}
\begin{centering}
\includegraphics[width=\columnwidth]{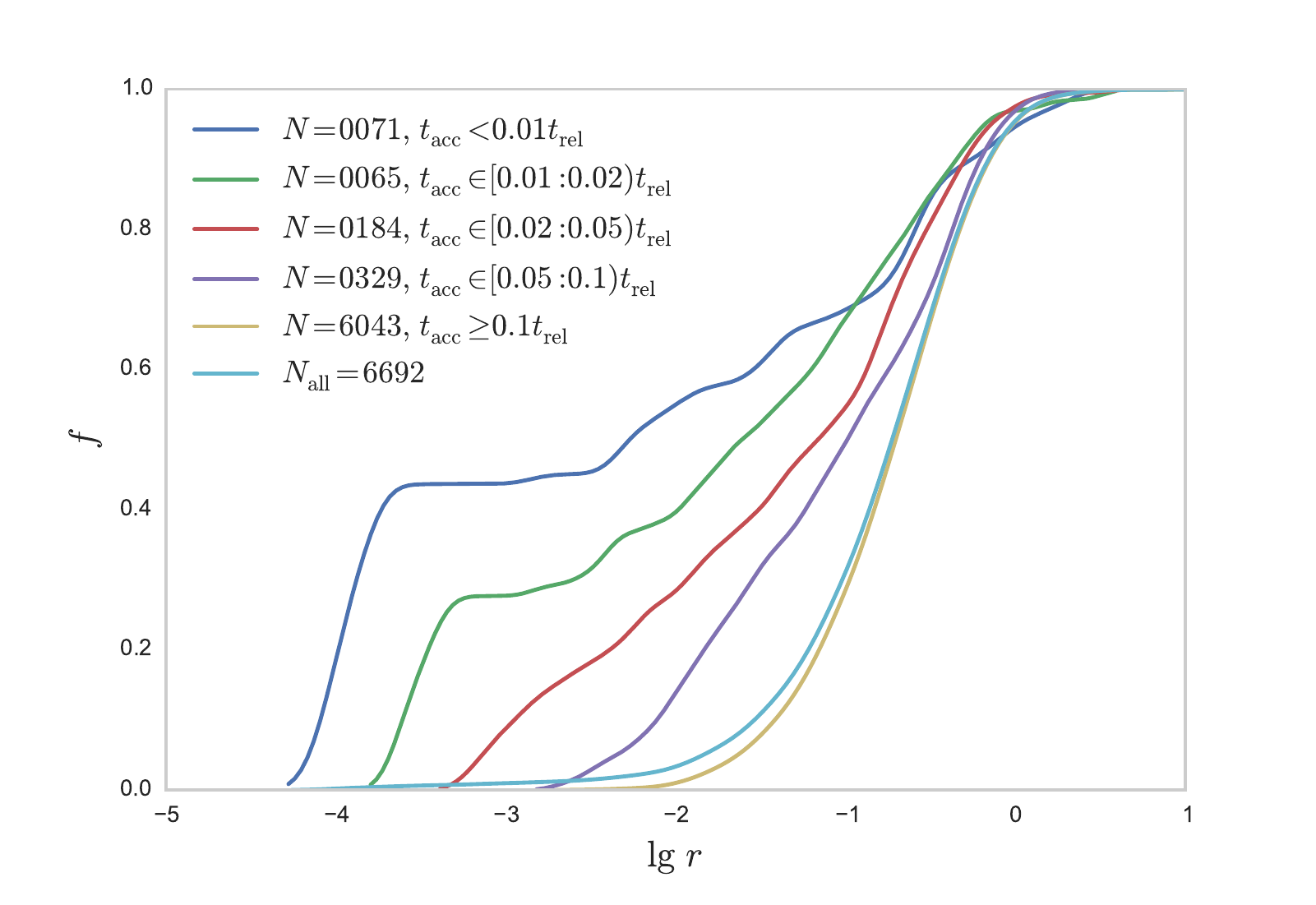}\\
\par\end{centering}
\caption{The cumulative distribution of the time intervals left to accretion at $t_\mathrm{cur} = 1.0\,t_\mathrm{rel}$ of all particles captured in in the time interval from 1.0 to 2.0\,$t_\mathrm{rel}$.}
\label{fig:longterm}
\end{figure}

The total number of accreted stars onto the SMBH is greater than the total number of stars captured by the AD because there is also a contribution from higher eccentricity orbits (plunge types 2 and 3 from Paper~II). Fig. \ref{fig:longterm} shows the long-term origin of accreted stars in the interval 1 - 2 $t_\mathrm{rel}$ including all plunge types. We clearly see that $\sim$50\% of stars accreted quickly are captured by the AD. On the other hand a significant fraction originate from $r=0.1-1$ and are first scattered into the loss cone before being accreted or captured. The change in the shapes of the curves shows the consistency with the derived value of the effective migration time.\\

We note that a closer look at the spatial distribution of the stellar disc particles reveals a small precession of the disc and warps. But the dynamics of individual NSD particles relative to each other is complex and lies beyond the scope of this paper.

\section{Scaling to real galactic centres}
\label{sec:scale}

\begin{table*}
\caption{Predicted nuclear stellar disc properties for a sample of galactic nuclei.}
\label{NSDtable}
\begin{tabular}{lllllll}
\hline
Object & $M_\mathrm{SMBH}$ & $r_\mathrm{inf}$ & $t_\mathrm{rel}$ & $M_\mathrm{NSD}$ & $R_\mathrm{NSD}$ & $T_\mathrm{migr}$\\
 & ($\msun$) & (pc) & (Gyr) & ($\msun$) & (pc) & (Gyr)\\
\hline
M 87       & $6.6 \times 10^{9}$ &  291 & $6 \times 10^{5}$ & $4.6 \times 10^{7}$ & 14.55 & 18000\\
NGC 3115   & $9.6 \times 10^{8}$ &   78 & $3.4 \times 10^{4}$ & $6.72 \times 10^{6}$ & 3.9 & 3180\\
NGC 4291   & $3.2 \times 10^{8}$ &   24 & 3400 & $2.24 \times 10^{6}$ & 1.2 & 258\\
M 31       & $1.5 \times 10^{8}$ &   25 & 2690 & $1.05 \times 10^{6}$ & 1.25 & 971\\
NGC 4486A  & $1.3 \times 10^{7}$ &  4.5 & 68.8 & $9.1 \times 10^{4}$ & 0.225 & 94\\
MW         & $4.0 \times 10^{6}$ &  1.4 & 7.2 & $2.8 \times 10^{4}$ & 0.07 & 9.4 \\
M 32       & $3.0 \times 10^{6}$ &  2.3 & 12.9 & $2.1 \times 10^{4}$ & 0.115 & 79.4\\
Circinus   & $1.7 \times 10^{6}$ &  0.29 & 2.3 & $1.2 \times 10^{4}$ & 0.015 & 0.6\\
\hline
\end{tabular}
\par\medskip
\begin{flushleft}\textbf{Notes.} We extrapolate the results to this sample of galactic nuclei (adopted from Paper~I and Paper~II). Columns 1--4 are the object's name, SMBH mass, influence radius (calculated from the stellar velocity dispersion) and half-mass relaxation time, respectively. Column 5 gives the mass of the NSD, column 6 gives the maximum radial size of the NSD, column 7 gives \textit{upper limits} of the `effective' migration time of a star through the AD to the SMBH.\end{flushleft}
\end{table*}

For calibration to real systems, the value of $Q_\mathrm{tot}$ has to be chosen accordingly (all other parameters in our simulations are independent of $N$). For example, when scaling the results of the simulation with $N = 1.28\times10^5$ to M87 using Eq.~\ref{Qtotdef}, Eq.~\ref{QtotNrel} and data from Table~\ref{NSDtable}, we get values of $Q_\mathrm{tot}(1.28\times10^5) = 5.42\times10^{-4}$ and $Q_\mathrm{tot}(6.6\times10^{10}) = 2.1\times10^{-9}$. Thus, the value is larger by 5 orders of magnitude due to the correct scaling in our simulation setting the dissipation and relaxation time-scales in correspondence (see Table 1 in Paper~I). 
For the other galaxies like the MW the mismatch of the used value for $Q_\mathrm{tot}$ is an issue, which needs to be discussed.
Since the accretion rate scales with the friction force $a_\mathrm{d} \sim Q_\mathrm{tot}\rho_\mathrm{g} \sim Q_\mathrm{tot}M_\mathrm{d}$ and time-scales scale with $t_\mathrm{diss} \sim 1/a_\mathrm{d} \sim 1/(Q_\mathrm{tot}M_d)$, for the Galactic centre with given $M_\mathrm{d} = 0.1 M_\mathrm{SMBH}$ the correct $Q_\mathrm{tot}$ would be $\sim 45$ times smaller leading to a 45 times smaller accretion rate and 45 times longer time to form the NSD. The mass of the NSD depends on the feeding by the friction force and the loss of stars by friction. If the scatter by 2-body relaxation is not important then the stationary stellar disc would have the same mass. If 2-body relaxation determines the feeding time-scale then the NSD  mass would be proportional to $\eta$ and would be larger by a factor of 45 in the Galactic centre if a stationary state is reached (at $T_\mathrm{migr}$).

Alternatively, the physical $Q_\mathrm{d} = 5$ per star from
ram pressure can be orders of magnitude larger due to dynamical friction
dependent on the relative velocities and the sound speed. Then the value of $Q_\mathrm{tot}$ in the simulations may be roughly consistent.
This can be the case in low mass SMBHs like in the MW and the Circinus galaxy, where the circular speed of the AD (and the velocity dispersion of the NSC stars) in the relevant distance range of the outer boundary of the NSD (=0.07\,pc for the MW) and the effective radius $R_\mathrm{eff}$ of the AD (see Paper~II; =0.2\,pc for the MW) falls below the escape speed at the surface of a solar-type star ($\approx 600\kms$, see Figure 1 in Paper~I).
For further calculations, we keep this scenario in mind. Note that \citet{Thun2016} have in great numerical detail  analysed how the `microscopic' value of $Q_\mathrm{d}$ depends on the relevant parameters (like the size of the moving body relative to the bow shock and the speed relative to the sound speed). They also provide useful analytic approximations for the supersonic case.

For the application to real galactic nuclei, we need to rescale all relevant quantities accordingly. For the total number of stars $N_\mathrm{real}$, which is greater than the number of particles in our simulation, $N_\mathrm{sim}$, one particle  in the simulation represents $N_\mathrm{scale} = N_\mathrm{real}/N_\mathrm{sim}$ stars. 
Given the core velocity dispersion $\sigma$ and mass of the SMBH (can be taken for example from \citealt{Gultekin2009}), one can calculate the influence radius $r_\mathrm{inf} =GM_\mathrm{SMBH}/\sigma^2$ and the relaxation time (given by Eq. 8 of Paper~II), taking into account that the half-mass radius equals $r_\mathrm{hm} = 3r_\mathrm{inf}$. The time and length scalings are done in a way that the relaxation time of the real system is the same as the relaxation time of the modelled system, as well as the influence radius of the SMBH in the real system is the same as in the modelled system (as described above, a more detailed description of the scaling procedure is given in Paper~I and Paper~II). We find the scaling factors $T_\mathrm{scale} = t_\mathrm{rel}^\mathrm{real}/t_\mathrm{rel}^\mathrm{sim}$ and $R_\mathrm{scale} = r_\mathrm{inf}^\mathrm{real}/r_\mathrm{inf}^\mathrm{sim}$. 

The capture rate by the AD for the MW is thus (see Table \ref{NSDtable}):
\begin{equation}
\frac{dN_\mathrm{cap}}{dt} = 3.65\times\frac{N_\mathrm{scale}}{T_\mathrm{scale}} \approx 140 [\mathrm{stars}\times \mathrm{Myr}^{-1}]  ,
\end{equation}
In other words, after 100 Myr of evolution, we expect 14000 stars to be trapped by a hypothetical gaseous disc, while most of them would be still in the migration phase. The `effective' migration time for the MW equals to $0.216$ Gyr (for the enhanced value of $Q_\mathrm{d}$ by dynamical friction). At $t=t_\mathrm{rel}$,  the mass of the NSD is 0.07\% of the initial total stellar mass of the NSC, converted to solar masses we get $M_\mathrm{NSD} \approx 3.0\times10^4 M_\mathrm{\odot}$. Note that this is an order of magnitude estimate, whereas in the real system due to mass segregation we expect more massive stars to populate the NSD. A detailed realistic simulation is our long-term goal. The mass of the NSD is of the order of magnitude of the observed mass of the young stellar disc(s) in the MW $\simeq10^4M_\mathrm{\odot}$ \citep{Bartko2010}, but the NSD stars should be older because of the long migration time. If the lifetime of the AD is shorter than the migration time than the mass of the NSD is expected to be smaller (e.g. by assuming a linear growth of the NSD).

Table \ref{NSDtable} gives the mass and size of the NSD as well as the migration time-scale, in physical units scaled according to the SMBH mass and its influence radius in nuclei of several nearby galaxies (adopted from Paper~I). This time-scales have to be treated as upper limits for the formation of the NSD, because only friction by ram pressure is taken into account. If the time is boosted by the dynamical friction, we can expect the presence of the NSD in lower mass systems while we do not expect stationary discs to form in massive galactic nuclei. 

The stellar migration time could be reduced by taking into account the pressure gradient of the AD to derive the circular speed of the gas, but this effect is very small and only relevant for the innermost particles.\\

\section{Summary and discussion} 
\label{sec:CON}

In this study we present the results from long-term simulations of a dense nuclear star cluster NSC surrounding a star-accreting SMBH and interacting with a central gaseous disc AD which acts as a drag force and dissipates stellar kinetic energy. First simulations of this kind were performed and described in Paper~I, improved in Paper~II. 
We examined the effect of star--disc interactions on the inner structure of the compact stellar cluster by means of direct $N$-body simulations. We found that the stars form a nuclear stellar disc NSD before being absorbed or disrupted by the SMBH. 
The AD leads to the formation of a NSD in very close vicinity of the SMBH with a mass of $M_\mathrm{NSD} \approx 0.007 M_\mathrm{SMBH}$. But the AD lifetime may be too short to fully form the NSD. We derived the effective stellar migration time through the AD towards the SMBH. Scaling the results to the Milky Way galaxy gives the mass of the NSD $M_\mathrm{NSD} \approx 3.0\times10^4 M_\mathrm{\odot}$ which is the same by order of magnitude as the observed disc of young massive stars in the MW,
but note that the NSD formed in our simulations consists of stars originating from
the old population of the spherical NSC. The outermost stars found in our NSD are located at
a distance of 0.07 pc from the SMBH. The observed young  Galactic Centre stellar disc resides
between 0.04 to 0.5 pc. We think that an NSD consisting of old stars, as found in our models,
could coexist with the observed stellar disc, but the old stars are just too faint to be detected close to the SMBH.
The second generation instrument for the Very Large Telescope Interferometer GRAVITY \citep{Eisenhauer2011} or the James Webb Space Telescope may be able to detect some of the NSD stars. Assuming that the young stars formed from the same AD (by disc fragmentation) which created the NSD (by trapping stars),
a detection of a disc of old stars (NSD) would
be a strong evidence for past activity and the former presence of an AGN disc and may give some hints on the efficiency of dynamical friction in the gaseous medium near the SMBH. 
 
Note that our results are nicely consistent with recent ideas about the non-stationary history of our
own Galactic centre, with sporadic AGN activity. The constant flow of gas to galactic nuclei inevitably
produces an accumulation of gas, the formation of a central disc. This will trigger both a central AGN 
flare-up activity as well as a central starburst after which the gaseous disc has disappeared \citep[e.g.][]{Novak2012}. A huge Fermi bubble has been detected on both sides
of the Galactic Centre \citep{Bordoloi2017}, which could be a remnant of an AGN evolutionary phase of
our own galaxy several Myr ago. 
  
The NSD is located inside the effective radius of the AD. As we have shown in Paper~II, it equals to $R_\mathrm{eff} = 0.032$ leading to an enclosed mass of $M_\mathrm{d}(<R_\mathrm{eff}) = 0.09 M_\mathrm{d}$ (note a typo: 42\% instead of the correct 9\% in Sec. 3.2 of Paper~II). That means that if we would cut the disc at $R_\mathrm{eff}$, the AD mass would be $\simeq1\%$ of the $M_\mathrm{SMBH}$ with essentially the same effect. If we reduce the surface density, the enclosed mass and the force would decrease proportionally leading to a smaller accretion rate. But the stellar disc mass would be similar, because it would just take longer to pass this phase. As a consequence the formation time would be larger. On the other hand, in Paper~I we have shown that dynamical friction would be very efficient if the relative velocity falls below the escape speed at the stellar surface ($\simeq600\kms$). The friction force would be orders of magnitude larger leading to the same accretion rate if we reduce the surface density of the AD accordingly. 
 
As was described in Sec.\ref{sec:method}, the contribution from dynamical friction (in other words gravitational focusing) was ignored in this study. But in case of subsonic motion the dissipation force may be enhanced leading to faster formation time of the NSD. This enhancement can be taken into account by replacing the drag coefficient $Q_\mathrm{d}$ by $Q_\mathrm{d} + (v_\mathrm{esc}/V_\mathrm{rel})^4\ln \Lambda$, where $v_\mathrm{esc}$ is the escape velocity from stellar surface, $V_\mathrm{rel}$ is the relative velocity of a star in the AD  and $\ln \Lambda\simeq10-20$ is the Coulomb logarithm. Given that for a SMBH mass of $\simeq10^8 M_\odot$, typical relative velocities are of the order of 1000$\kms$ at distances below 1pc, dynamical friction can be ignored for normal stars. But it can be sufficient for compact objects since the escape velocity from their surface is high. The supersonic motion of stars through the gaseous medium is an active field of research in astrophysics \cite[e.g.][]{Thun2016} and we will incorporate new results from this field to our simulations, such as $Q_\mathrm{d}$ as a function of $\rho_\mathrm{g}$ and $V_\mathrm{rel}$. 

It is more likely that NSDs may reside in low mass galactic nuclei (with $M_\mathrm{SMBH}\simeq 10^6 M_\odot$). We assumed long-lived ADs (several hundred Myr) but in reality, the AD may be short-lived and the NSD would not form completely within the lifetime of the AD. The observational estimates of AGN lifetimes give a wide range of values. For example, the AGN fraction in SDSS data implies long lifetimes of $t_\mathrm{AGN} > 10^8 \mathrm{yr}$ \citep{MillerEtAl2003}, but \citet{Schawinski2015} argue that the SMBH growth period consists of many short episodes of activity with $t_\mathrm{AGN} \simeq10^5\mathrm{yr}$. The estimates based on the effects of quasar proximity on the surrounding gas (studying the absorption lines) yield AGN lifetimes of order of $10^6-10^7$ years \citep[e.g.][and references therein]{Schirber2004, Syphers2014, Segers2017}. It is likely that a NSD forms even in the case of a short active phase, if AGN activity repeats and a gaseous disc forms in the same orientation in each such short episode \citep{Schawinski2015}. Our results show that for shorter disc lifetimes (see Fig. \ref{fig:ncap}) we will get a NSD with smaller mass. We find that after around $10^9$ years a stationary state is established, if the disc lives that long. 
The migration time for the low SMBH mass galaxies ($\simeq 10^6 M_{\odot} )$ is comparable with the upper limits on the $t_\mathrm{AGN}$ making this objects the best candidates for the presence of a NSD. One of the nearby active galaxies Circinus (added in Table \ref{NSDtable}) features a NSD formation time of 60 Myr and potentially may host a NSD in its nucleus. On the other hand for the high mass SMBH galaxies such as M87, the NSD formation time-scale is too large to produce a stationary stellar disc.

If the AD disappears when the NSD is already formed then the latter will survive for a fraction of relaxation time. Although the orbital orientations of stars in the NSD may be randomised by resonant relaxation \citep[e.g.][]{Hopman2006}, this mechanism was proposed to explain random orbital orientations of the Galactic Centre S-stars \citep[e.g.][]{Perets2009a,Antonini&Merritt2013}. But whether the resonant relaxation is really dominant in real systems, with mass spectrum, small deviations from spherical symmetry (like our disc potential) are highly controversial. We did not take into account the self-gravity of the AD and the impact of the NSC potential to the rotation speed of the AD.
Inside $R_\mathrm{eff}$ the flattening of the potential due to the AD is already a factor of 10 smaller and with the smaller surface density combined with dynamical friction, it would be completely negligible. The outer radius of the AD is chosen to be equal to the influence radius of the SMBH in the sense of $M_\mathrm{cl}(<R_\mathrm{d}) = M_\mathrm{SMBH}$, therefore the AD mass is 10\% of the $M_\mathrm{SMBH}$ and 5\% of the $M_\mathrm{SMBH}+M_\mathrm{cl}$ at $R_\mathrm{d}$. The correction to the rotation curve would be dominated by self-gravity of the cluster. 

\citet{MK2005} argue that stars captured by the accretion disc are eventually destroyed and their matter diffused within the AD. This might happen in the very inner region of the galactic nucleus where contact stellar collisions play an important role, but our simulations do not resolve to that extent and the NSD forms further outside. The evolved and more massive stars have lower surface densities and their interaction with the AD can strip outer layers of the crossing star resulting in shallower stellar density profile \citep{Amaro-Seoane2014, Kieffer2016}.  
 
The star--disc interactions in AGN with a stellar mass spectrum and stellar evolution are planned to be examined in future work. In particular, the formation, evolution and subsequent merging of binary black holes in the gaseous disc are of great interest. While residing in the AD, black holes can accrete material and merge with masses comparable to those detected by LIGO \citep{Abbott2016}. \cite{Bartos2017} and \cite{McKernan2017} used semi-analytic approaches to calculate the detection rate of such events by LIGO, but their estimates span three orders of magnitude \citep{McKernan2017}. \citet{Haggard2010} found that $0.16\%\pm0.06\%$ of all galaxies in the local universe are active (some nearby active galaxies are NGC4051 at a distance of $\simeq\mathrm{10\,Mpc}$ and NGC4151 at $\simeq\mathrm{14\,Mpc}$; \citealt{Bentz2015}), implying a very large number of AGN ADs within the LIGO sensitivity volume. High resolution direct $N$-body simulations including realistic physics of the gaseous ADs may set much better constrains on this problem. Exploring effects of stellar crossing on the gaseous disc requires detailed SPH or hydrodynamical simulations of the AD including all relevant physics. 
A fully realistic direct $N$-body simulation of AGNs including gas physics in the AD remains as our long-term goal.

\section{Acknowledgements}

TP and BS acknowledge the support within program N0003-5/PCF-15-AKMIR by the Ministry of Education and Science of the Republic of Kazakhstan. YM was supported by the European Research Council under the European Union's Horizon 2020 Programme, ERC-2014-STG grant GalNUC 638435.
This work has benefited very much from funding of exchange and collaboration between Germany and Kazakhstan by Volkswagen Foundation under the project `STARDISK -- Simulating Dense Star--Gas Systems in Galactic Nuclei using special hardware' (I/81 396). This work has been partly supported by the project ``GRACE 2: Scientific simulations using programmable hardware" of the Volkswagen Foundation (grants I84678/84680), especially through the use of the kepler GPU cluster at ARI.

We acknowledge the support by Chinese Academy of Sciences through the
Silk Road Project at NAOC, through the ``Qianren" special foreign experts
program and also under the President's International Fellowship for Visiting
Scientists program of CAS. We thank the Excellence Initiative at the University of Heidelberg,
which supported us by measures for international research collaborations

P.B., and R.S. acknowledge also the Strategic Priority Research Program*
(Pilot B)* ``Multi-wavelength gravitational wave universe" of the Chinese
Academy of Sciences (No. XDB23040100).

PB acknowledge the support of the Volkswagen Foundation under the Trilateral
Partnerships grant No. 90411 and the special support by the NASU under
the Main Astronomical Observatory GRID/GPU computing cluster project.

The special GPU accelerated supercomputer laohu at the Centre of Information and Computing at National Astronomical Observatories, Chinese Academy of Sciences, funded by Ministry of Finance of Peoples Republic of China under the grant ZDYZ2008-2, has been used for some of the largest simulations. The authors gratefully acknowledge the computing time granted by the John von Neumann Institute for Computing (NIC)
and provided on the supercomputer JURECA \citep{jureca} at J\"ulich Supercomputing Centre (JSC) through grant HHD28. 

We thank H. Perets for his useful suggestions and comments as referee. 

\bibliography{SDP3Refs}

\begin{thebibliography}{}
\makeatletter
\relax
\def\mn@urlcharsother{\let\do\@makeother \do\$\do\&\do\#\do\^\do\_\do\%\do\~}
\def\mn@doi{\begingroup\mn@urlcharsother \@ifnextchar [ {\mn@doi@}
  {\mn@doi@[]}}
\def\mn@doi@[#1]#2{\def\@tempa{#1}\ifx\@tempa\@empty \href
  {http://dx.doi.org/#2} {doi:#2}\else \href {http://dx.doi.org/#2} {#1}\fi
  \endgroup}
\def\mn@eprint#1#2{\mn@eprint@#1:#2::\@nil}
\def\mn@eprint@arXiv#1{\href {http://arxiv.org/abs/#1} {{\tt arXiv:#1}}}
\def\mn@eprint@dblp#1{\href {http://dblp.uni-trier.de/rec/bibtex/#1.xml}
  {dblp:#1}}
\def\mn@eprint@#1:#2:#3:#4\@nil{\def\@tempa {#1}\def\@tempb {#2}\def\@tempc
  {#3}\ifx \@tempc \@empty \let \@tempc \@tempb \let \@tempb \@tempa \fi \ifx
  \@tempb \@empty \def\@tempb {arXiv}\fi \@ifundefined
  {mn@eprint@\@tempb}{\@tempb:\@tempc}{\expandafter \expandafter \csname
  mn@eprint@\@tempb\endcsname \expandafter{\@tempc}}}

\bibitem[\protect\citeauthoryear{{Abbott} et~al.,}{{Abbott}
  et~al.}{2016}]{Abbott2016}
{Abbott} B.~P.,  et~al., 2016, \mn@doi [Phys. Rev. Lett]
  {10.1103/PhysRevLett.116.061102}, \href
  {http://adsabs.harvard.edu/abs/2016PhRvL.116f1102A} {116, 061102}

\bibitem[\protect\citeauthoryear{{Amaro-Seoane} \& {Chen}}{{Amaro-Seoane} \&
  {Chen}}{2014}]{Amaro-Seoane2014}
{Amaro-Seoane} P.,  {Chen} X.,  2014, \mn@doi [\apjl]
  {10.1088/2041-8205/781/1/L18}, \href
  {http://adsabs.harvard.edu/abs/2014ApJ...781L..18A} {781, L18}

\bibitem[\protect\citeauthoryear{{Antonini} \& {Merritt}}{{Antonini} \&
  {Merritt}}{2013}]{Antonini&Merritt2013}
{Antonini} F.,  {Merritt} D.,  2013, \mn@doi [\apjl]
  {10.1088/2041-8205/763/1/L10}, \href
  {http://adsabs.harvard.edu/abs/2013ApJ...763L..10A} {763, L10}

\bibitem[\protect\citeauthoryear{{Antonucci}}{{Antonucci}}{1993}]{Antonucci1993}
{Antonucci} R.,  1993, \mn@doi [\araa] {10.1146/annurev.aa.31.090193.002353},
  \href {http://adsabs.harvard.edu/abs/1993ARA%26A..31..473A} {31, 473}

\bibitem[\protect\citeauthoryear{{Artymowicz}, {Lin}  \&
  {Wampler}}{{Artymowicz} et~al.}{1993}]{ArtymowiczEtAl1993}
{Artymowicz} P.,  {Lin} D.~N.~C.,   {Wampler} E.~J.,  1993, \mn@doi [ApJ]
  {10.1086/172690}, \href {http://adsabs.harvard.edu/abs/1993ApJ...409..592A}
  {409, 592}

\bibitem[\protect\citeauthoryear{{Bartko} et~al.,}{{Bartko}
  et~al.}{2009}]{Bartko2009}
{Bartko} H.,  et~al., 2009, \mn@doi [\apj] {10.1088/0004-637X/697/2/1741},
  \href {http://adsabs.harvard.edu/abs/2009ApJ...697.1741B} {697, 1741}

\bibitem[\protect\citeauthoryear{{Bartko} et~al.,}{{Bartko}
  et~al.}{2010}]{Bartko2010}
{Bartko} H.,  et~al., 2010, \mn@doi [\apj] {10.1088/0004-637X/708/1/834}, \href
  {http://adsabs.harvard.edu/abs/2010ApJ...708..834B} {708, 834}

\bibitem[\protect\citeauthoryear{{Bartos}, {Kocsis}, {Haiman}  \&
  {M{\'a}rka}}{{Bartos} et~al.}{2017}]{Bartos2017}
{Bartos} I.,  {Kocsis} B.,  {Haiman} Z.,   {M{\'a}rka} S.,  2017, \mn@doi
  [\apj] {10.3847/1538-4357/835/2/165}, \href
  {http://adsabs.harvard.edu/abs/2017ApJ...835..165B} {835, 165}

\bibitem[\protect\citeauthoryear{{Baruteau}, {Cuadra}  \& {Lin}}{{Baruteau}
  et~al.}{2011}]{Baruteau2011}
{Baruteau} C.,  {Cuadra} J.,   {Lin} D.~N.~C.,  2011, \mn@doi [\apj]
  {10.1088/0004-637X/726/1/28}, \href
  {http://adsabs.harvard.edu/abs/2011ApJ...726...28B} {726, 28}

\bibitem[\protect\citeauthoryear{{Bellovary}, {Mac Low}, {McKernan}  \&
  {Ford}}{{Bellovary} et~al.}{2016}]{Bellovary2016}
{Bellovary} J.~M.,  {Mac Low} M.-M.,  {McKernan} B.,   {Ford} K.~E.~S.,  2016,
  \mn@doi [\apjl] {10.3847/2041-8205/819/2/L17}, \href
  {http://adsabs.harvard.edu/abs/2016ApJ...819L..17B} {819, L17}

\bibitem[\protect\citeauthoryear{{Bentz} \& {Katz}}{{Bentz} \&
  {Katz}}{2015}]{Bentz2015}
{Bentz} M.~C.,  {Katz} S.,  2015, \mn@doi [\pasp] {10.1086/679601}, \href
  {http://adsabs.harvard.edu/abs/2015PASP..127...67B} {127, 67}

\bibitem[\protect\citeauthoryear{{Bordoloi} et~al.,}{{Bordoloi}
  et~al.}{2017}]{Bordoloi2017}
{Bordoloi} R.,  et~al., 2017, \mn@doi [ApJ] {10.3847/1538-4357/834/2/191},
  \href {http://cdsads.u-strasbg.fr/abs/2017ApJ...834..191B} {834, 191}

\bibitem[\protect\citeauthoryear{{Buchholz}, {Sch{\"o}del}  \&
  {Eckart}}{{Buchholz} et~al.}{2009}]{Buchholz2009}
{Buchholz} R.~M.,  {Sch{\"o}del} R.,   {Eckart} A.,  2009, \mn@doi [\aap]
  {10.1051/0004-6361/200811497}, \href
  {http://adsabs.harvard.edu/abs/2009A%26A...499..483B} {499, 483}

\bibitem[\protect\citeauthoryear{{Courant} \& {Friedrichs}}{{Courant} \&
  {Friedrichs}}{1948}]{CourantFriedrichs1948}
{Courant} R.,  {Friedrichs} K.~O.,  1948, {Supersonic flow and shock waves}

\bibitem[\protect\citeauthoryear{{Eisenhauer}, {Perrin}, {Brandner},
  {Straubmeier}, {Perraut}, {Amorim}, {Sch{\"o}ller}  \&
  {Gillessen}}{{Eisenhauer} et~al.}{2011}]{Eisenhauer2011}
{Eisenhauer} F.,  {Perrin} G.,  {Brandner} W.,  {Straubmeier} C.,  {Perraut}
  K.,  {Amorim} A.,  {Sch{\"o}ller} M.,   {Gillessen} S. e.~a.,  2011, The
  Messenger, \href {http://adsabs.harvard.edu/abs/2011Msngr.143...16E} {143,
  16}

\bibitem[\protect\citeauthoryear{{Gallego-Cano}, {Sch{\"o}del}, {Dong},
  {Nogueras-Lara}, {Gallego-Calvente}, {Amaro-Seoane}  \&
  {Baumgardt}}{{Gallego-Cano} et~al.}{2017}]{Gallego-Cano2017}
{Gallego-Cano} E.,  {Sch{\"o}del} R.,  {Dong} H.,  {Nogueras-Lara} F.,
  {Gallego-Calvente} A.~T.,  {Amaro-Seoane} P.,   {Baumgardt} H.,  2017,
  preprint, \href {http://adsabs.harvard.edu/abs/2017arXiv170103816G} {}
  (\mn@eprint {arXiv} {1701.03816})

\bibitem[\protect\citeauthoryear{{Genzel}, {Eisenhauer}  \&
  {Gillessen}}{{Genzel} et~al.}{2010}]{Genzel2010}
{Genzel} R.,  {Eisenhauer} F.,   {Gillessen} S.,  2010, \mn@doi [Rev. Mod.
  Phys.] {10.1103/RevModPhys.82.3121}, \href
  {http://adsabs.harvard.edu/abs/2010RvMP...82.3121G} {82, 3121}

\bibitem[\protect\citeauthoryear{{Gillessen} et~al.,}{{Gillessen}
  et~al.}{2017}]{Gillessen2017}
{Gillessen} S.,  et~al., 2017, \mn@doi [\apj] {10.3847/1538-4357/aa5c41}, \href
  {http://adsabs.harvard.edu/abs/2017ApJ...837...30G} {837, 30}

\bibitem[\protect\citeauthoryear{{G{\"u}ltekin} et~al.,}{{G{\"u}ltekin}
  et~al.}{2009}]{Gultekin2009}
{G{\"u}ltekin} K.,  et~al., 2009, \mn@doi [\apj] {10.1088/0004-637X/698/1/198},
  \href {http://adsabs.harvard.edu/abs/2009ApJ...698..198G} {698, 198}

\bibitem[\protect\citeauthoryear{{Haggard}, {Green}, {Anderson}, {Constantin},
  {Aldcroft}, {Kim}  \& {Barkhouse}}{{Haggard} et~al.}{2010}]{Haggard2010}
{Haggard} D.,  {Green} P.~J.,  {Anderson} S.~F.,  {Constantin} A.,  {Aldcroft}
  T.~L.,  {Kim} D.-W.,   {Barkhouse} W.~A.,  2010, \mn@doi [ApJ]
  {10.1088/0004-637X/723/2/1447}, \href
  {http://adsabs.harvard.edu/abs/2010ApJ...723.1447H} {723, 1447}

\bibitem[\protect\citeauthoryear{{Harfst}, {Gualandris}, {Merritt}, {Spurzem},
  {Portegies Zwart}  \& {Berczik}}{{Harfst} et~al.}{2007}]{HarfstEtAl2007}
{Harfst} S.,  {Gualandris} A.,  {Merritt} D.,  {Spurzem} R.,  {Portegies Zwart}
  S.,   {Berczik} P.,  2007, \mn@doi [\na] {10.1016/j.newast.2006.11.003},
  \href {http://adsabs.harvard.edu/abs/2007NewA...12..357H} {12, 357}

\bibitem[\protect\citeauthoryear{{Hopman} \& {Alexander}}{{Hopman} \&
  {Alexander}}{2006}]{Hopman2006}
{Hopman} C.,  {Alexander} T.,  2006, \mn@doi [\apj] {10.1086/504400}, \href
  {http://adsabs.harvard.edu/abs/2006ApJ...645.1152H} {645, 1152}

\bibitem[\protect\citeauthoryear{{J\"{u}lich Supercomputing
  Centre}}{{J\"{u}lich Supercomputing Centre}}{2016}]{jureca}
{J\"{u}lich Supercomputing Centre} 2016, \mn@doi [Journal of large-scale
  research facilities] {10.17815/jlsrf-2-121}, 2

\bibitem[\protect\citeauthoryear{{Just}, {Yurin}, {Makukov}, {Berczik},
  {Omarov}, {Spurzem}  \& {Vilkoviskij}}{{Just} et~al.}{2012}]{JustEtAl2012}
{Just} A.,  {Yurin} D.,  {Makukov} M.,  {Berczik} P.,  {Omarov} C.,  {Spurzem}
  R.,   {Vilkoviskij} E.~Y.,  2012, \mn@doi [ApJ] {10.1088/0004-637X/758/1/51},
  \href {http://adsabs.harvard.edu/abs/2012ApJ...758...51J} {758, 51}

\bibitem[\protect\citeauthoryear{{Kennedy}, {Meiron}, {Shukirgaliyev},
  {Panamarev}, {Berczik}, {Just}  \& {Spurzem}}{{Kennedy}
  et~al.}{2016}]{KenEtAl2016}
{Kennedy} G.~F.,  {Meiron} Y.,  {Shukirgaliyev} B.,  {Panamarev} T.,  {Berczik}
  P.,  {Just} A.,   {Spurzem} R.,  2016, \mn@doi [\mnras]
  {10.1093/mnras/stw908}, \href
  {http://adsabs.harvard.edu/abs/2016MNRAS.460..240K} {460, 240}

\bibitem[\protect\citeauthoryear{{Kieffer} \& {Bogdanovi{\'c}}}{{Kieffer} \&
  {Bogdanovi{\'c}}}{2016}]{Kieffer2016}
{Kieffer} T.~F.,  {Bogdanovi{\'c}} T.,  2016, \mn@doi [\apj]
  {10.3847/0004-637X/823/2/155}, \href
  {http://adsabs.harvard.edu/abs/2016ApJ...823..155K} {823, 155}

\bibitem[\protect\citeauthoryear{{Kocsis} \& {Tremaine}}{{Kocsis} \&
  {Tremaine}}{2011}]{Kocsis2011}
{Kocsis} B.,  {Tremaine} S.,  2011, \mn@doi [\mnras]
  {10.1111/j.1365-2966.2010.17897.x}, \href
  {http://adsabs.harvard.edu/abs/2011MNRAS.412..187K} {412, 187}

\bibitem[\protect\citeauthoryear{{Leigh}, {Mastrobuono-Battisti}, {Perets}  \&
  {B{\"o}ker}}{{Leigh} et~al.}{2014}]{Leigh2014}
{Leigh} N.~W.~C.,  {Mastrobuono-Battisti} A.,  {Perets} H.~B.,   {B{\"o}ker}
  T.,  2014, \mn@doi [\mnras] {10.1093/mnras/stu622}, \href
  {http://adsabs.harvard.edu/abs/2014MNRAS.441..919L} {441, 919}

\bibitem[\protect\citeauthoryear{{Levin} \& {Beloborodov}}{{Levin} \&
  {Beloborodov}}{2003}]{LevinBeloborodov2003}
{Levin} Y.,  {Beloborodov} A.~M.,  2003, \mn@doi [ApJL] {10.1086/376675}, \href
  {http://adsabs.harvard.edu/abs/2003ApJ...590L..33L} {590, L33}

\bibitem[\protect\citeauthoryear{{Li}, {Liu}, {Berczik}  \& {Spurzem}}{{Li}
  et~al.}{2017}]{Li2017}
{Li} S.,  {Liu} F.~K.,  {Berczik} P.,   {Spurzem} R.,  2017, \mn@doi [\apj]
  {10.3847/1538-4357/834/2/195}, \href
  {http://adsabs.harvard.edu/abs/2017ApJ...834..195L} {834, 195}

\bibitem[\protect\citeauthoryear{{Lu}, {Ghez}, {Hornstein}, {Morris}, {Becklin}
   \& {Matthews}}{{Lu} et~al.}{2009}]{Lu2009}
{Lu} J.~R.,  {Ghez} A.~M.,  {Hornstein} S.~D.,  {Morris} M.~R.,  {Becklin}
  E.~E.,   {Matthews} K.,  2009, \mn@doi [ApJ] {10.1088/0004-637X/690/2/1463},
  \href {http://adsabs.harvard.edu/abs/2009ApJ...690.1463L} {690, 1463}

\bibitem[\protect\citeauthoryear{{McKernan}, {Ford}, {Lyra}, {Perets}, {Winter}
   \& {Yaqoob}}{{McKernan} et~al.}{2011}]{Mckernan2011}
{McKernan} B.,  {Ford} K.~E.~S.,  {Lyra} W.,  {Perets} H.~B.,  {Winter} L.~M.,
   {Yaqoob} T.,  2011, \mn@doi [\mnras] {10.1111/j.1745-3933.2011.01132.x},
  \href {http://adsabs.harvard.edu/abs/2011MNRAS.417L.103M} {417, L103}

\bibitem[\protect\citeauthoryear{{McKernan} et~al.,}{{McKernan}
  et~al.}{2017}]{McKernan2017}
{McKernan} B.,  et~al., 2017, preprint, \href
  {http://adsabs.harvard.edu/abs/2017arXiv170207818M} {} (\mn@eprint {arXiv}
  {1702.07818})

\bibitem[\protect\citeauthoryear{{Miller}, {Nichol}, {G{\'o}mez}, {Hopkins}  \&
  {Bernardi}}{{Miller} et~al.}{2003}]{MillerEtAl2003}
{Miller} C.~J.,  {Nichol} R.~C.,  {G{\'o}mez} P.~L.,  {Hopkins} A.~M.,
  {Bernardi} M.,  2003, \mn@doi [ApJ] {10.1086/378383}, \href
  {http://adsabs.harvard.edu/abs/2003ApJ...597..142M} {597, 142}

\bibitem[\protect\citeauthoryear{{Miralda-Escud{\'e}} \&
  {Kollmeier}}{{Miralda-Escud{\'e}} \& {Kollmeier}}{2005}]{MK2005}
{Miralda-Escud{\'e}} J.,  {Kollmeier} J.~A.,  2005, \mn@doi [ApJ]
  {10.1086/426467}, \href {http://adsabs.harvard.edu/abs/2005ApJ...619...30M}
  {619, 30}

\bibitem[\protect\citeauthoryear{{Nayakshin} \& {Cuadra}}{{Nayakshin} \&
  {Cuadra}}{2005}]{Nayakshin2005}
{Nayakshin} S.,  {Cuadra} J.,  2005, \mn@doi [\aap]
  {10.1051/0004-6361:20042052}, \href
  {http://adsabs.harvard.edu/abs/2005A%26A...437..437N} {437, 437}

\bibitem[\protect\citeauthoryear{{Nayakshin}, {Cuadra}  \&
  {Springel}}{{Nayakshin} et~al.}{2007}]{Nayakshin2007}
{Nayakshin} S.,  {Cuadra} J.,   {Springel} V.,  2007, \mn@doi [\mnras]
  {10.1111/j.1365-2966.2007.11938.x}, \href
  {http://adsabs.harvard.edu/abs/2007MNRAS.379...21N} {379, 21}

\bibitem[\protect\citeauthoryear{{Netzer}}{{Netzer}}{2015}]{Netzer2015}
{Netzer} H.,  2015, \mn@doi [\araa] {10.1146/annurev-astro-082214-122302},
  \href {http://adsabs.harvard.edu/abs/2015ARA%26A..53..365N} {53, 365}

\bibitem[\protect\citeauthoryear{{Novak}, {Ostriker}  \& {Ciotti}}{{Novak}
  et~al.}{2012}]{Novak2012}
{Novak} G.~S.,  {Ostriker} J.~P.,   {Ciotti} L.,  2012, \mn@doi [MNRAS]
  {10.1111/j.1365-2966.2012.21844.x}, \href
  {http://cdsads.u-strasbg.fr/abs/2012MNRAS.427.2734N} {427, 2734}

\bibitem[\protect\citeauthoryear{{Novikov} \& {Thorne}}{{Novikov} \&
  {Thorne}}{1973}]{NT1973}
{Novikov} I.~D.,  {Thorne} K.~S.,  1973, in {Dewitt} C.,  {Dewitt} B.~S.,  eds,
  Black Holes (Les Astres Occlus). pp 343--450

\bibitem[\protect\citeauthoryear{{Ostriker}}{{Ostriker}}{1999}]{Ostriker1999}
{Ostriker} E.~C.,  1999, \mn@doi [ApJ] {10.1086/306858}, \href
  {http://adsabs.harvard.edu/abs/1999ApJ...513..252O} {513, 252}

\bibitem[\protect\citeauthoryear{{Paumard} et~al.,}{{Paumard}
  et~al.}{2006}]{PaumardEtAl2006}
{Paumard} T.,  et~al., 2006, \mn@doi [ApJ] {10.1086/503273}, \href
  {http://adsabs.harvard.edu/abs/2006ApJ...643.1011P} {643, 1011}

\bibitem[\protect\citeauthoryear{{Perets}, {Gualandris}, {Kupi}, {Merritt}  \&
  {Alexander}}{{Perets} et~al.}{2009}]{Perets2009a}
{Perets} H.~B.,  {Gualandris} A.,  {Kupi} G.,  {Merritt} D.,   {Alexander} T.,
  2009, \mn@doi [\apj] {10.1088/0004-637X/702/2/884}, \href
  {http://adsabs.harvard.edu/abs/2009ApJ...702..884P} {702, 884}

\bibitem[\protect\citeauthoryear{{Rauch}}{{Rauch}}{1995}]{Rauch1995}
{Rauch} K.~P.,  1995, MNRAS, \href
  {http://adsabs.harvard.edu/abs/1995MNRAS.275..628R} {275, 628}

\bibitem[\protect\citeauthoryear{{Rauch}}{{Rauch}}{1999}]{Rauch1999}
{Rauch} K.~P.,  1999, \mn@doi [ApJ] {10.1086/306953}, \href
  {http://adsabs.harvard.edu/abs/1999ApJ...514..725R} {514, 725}

\bibitem[\protect\citeauthoryear{{Schawinski}, {Koss}, {Berney}  \&
  {Sartori}}{{Schawinski} et~al.}{2015}]{Schawinski2015}
{Schawinski} K.,  {Koss} M.,  {Berney} S.,   {Sartori} L.~F.,  2015, \mn@doi
  [\mnras] {10.1093/mnras/stv1136}, \href
  {http://adsabs.harvard.edu/abs/2015MNRAS.451.2517S} {451, 2517}

\bibitem[\protect\citeauthoryear{{Schirber}, {Miralda-Escud{\'e}}  \&
  {McDonald}}{{Schirber} et~al.}{2004}]{Schirber2004}
{Schirber} M.,  {Miralda-Escud{\'e}} J.,   {McDonald} P.,  2004, \mn@doi [\apj]
  {10.1086/421451}, \href {http://adsabs.harvard.edu/abs/2004ApJ...610..105S}
  {610, 105}

\bibitem[\protect\citeauthoryear{{Segers}, {Oppenheimer}, {Schaye}  \&
  {Richings}}{{Segers} et~al.}{2017}]{Segers2017}
{Segers} M.~C.,  {Oppenheimer} B.~D.,  {Schaye} J.,   {Richings} A.~J.,  2017,
  \mn@doi [\mnras] {10.1093/mnras/stx1633}, \href
  {http://adsabs.harvard.edu/abs/2017MNRAS.471.1026S} {471, 1026}

\bibitem[\protect\citeauthoryear{{Shakura} \& {Sunyaev}}{{Shakura} \&
  {Sunyaev}}{1973}]{SS1973}
{Shakura} N.~I.,  {Sunyaev} R.~A.,  1973, \aap, \href
  {http://adsabs.harvard.edu/abs/1973A%26A....24..337S} {24, 337}

\bibitem[\protect\citeauthoryear{{Shukirgaliyev}}{{Shukirgaliyev}}{2016}]{Shukirgaliyev2016}
{Shukirgaliyev} B.,  2016, in {Meiron} Y.,  {Li} S.,  {Liu} F.-K.,   {Spurzem}
  R.,  eds,  IAU Symposium Vol. 312, Star Clusters and Black Holes in Galaxies
  across Cosmic Time. pp 113--117 (\mn@eprint {arXiv} {1610.02838}),
  \mn@doi{10.1017/S1743921315007644}

\bibitem[\protect\citeauthoryear{{Stone}, {Metzger}  \& {Haiman}}{{Stone}
  et~al.}{2017}]{Stone2017}
{Stone} N.~C.,  {Metzger} B.~D.,   {Haiman} Z.,  2017, \mn@doi [\mnras]
  {10.1093/mnras/stw2260}, \href
  {http://adsabs.harvard.edu/abs/2017MNRAS.464..946S} {464, 946}

\bibitem[\protect\citeauthoryear{{Syphers} \& {Shull}}{{Syphers} \&
  {Shull}}{2014}]{Syphers2014}
{Syphers} D.,  {Shull} J.~M.,  2014, \mn@doi [\apj]
  {10.1088/0004-637X/784/1/42}, \href
  {http://adsabs.harvard.edu/abs/2014ApJ...784...42S} {784, 42}

\bibitem[\protect\citeauthoryear{{Thun}, {Kuiper}, {Schmidt}  \& {Kley}}{{Thun}
  et~al.}{2016}]{Thun2016}
{Thun} D.,  {Kuiper} R.,  {Schmidt} F.,   {Kley} W.,  2016, \mn@doi [\aap]
  {10.1051/0004-6361/201527629}, \href
  {http://adsabs.harvard.edu/abs/2016A%26A...589A..10T} {589, A10}

\bibitem[\protect\citeauthoryear{{Urry} \& {Padovani}}{{Urry} \&
  {Padovani}}{1995}]{Urry1995}
{Urry} C.~M.,  {Padovani} P.,  1995, \mn@doi [\pasp] {10.1086/133630}, \href
  {http://adsabs.harvard.edu/abs/1995PASP..107..803U} {107, 803}

\bibitem[\protect\citeauthoryear{{Vilkoviskij} \& {Czerny}}{{Vilkoviskij} \&
  {Czerny}}{2002}]{VilkoviskijCzerny2002}
{Vilkoviskij} E.~Y.,  {Czerny} B.,  2002, \mn@doi [\aap]
  {10.1051/0004-6361:20020255}, \href
  {http://adsabs.harvard.edu/abs/2002A%26A...387..804V} {387, 804}

\bibitem[\protect\citeauthoryear{{Yelda}, {Ghez}, {Lu}, {Do}, {Meyer}, {Morris}
   \& {Matthews}}{{Yelda} et~al.}{2014}]{Yelda2014}
{Yelda} S.,  {Ghez} A.~M.,  {Lu} J.~R.,  {Do} T.,  {Meyer} L.,  {Morris} M.~R.,
    {Matthews} K.,  2014, \mn@doi [\apj] {10.1088/0004-637X/783/2/131}, \href
  {http://adsabs.harvard.edu/abs/2014ApJ...783..131Y} {783, 131}

\bibitem[\protect\citeauthoryear{{Zhong}, {Berczik}  \& {Spurzem}}{{Zhong}
  et~al.}{2014}]{Zhong2014}
{Zhong} S.,  {Berczik} P.,   {Spurzem} R.,  2014, \mn@doi [\apj]
  {10.1088/0004-637X/792/2/137}, \href
  {http://adsabs.harvard.edu/abs/2014ApJ...792..137Z} {792, 137}

\bibitem[\protect\citeauthoryear{{Zhong}, {Berczik}  \& {Spurzem}}{{Zhong}
  et~al.}{2015}]{Zhong2015}
{Zhong} S.,  {Berczik} P.,   {Spurzem} R.,  2015, \mn@doi [\apj]
  {10.1088/0004-637X/811/1/22}, \href
  {http://adsabs.harvard.edu/abs/2015ApJ...811...22Z} {811, 22}

\makeatother
\end{thebibliography}

\end{document}